\documentclass{aa}
\usepackage{graphicx, amsmath, color, mathrsfs}
\usepackage[normalem]{ ulem }
\usepackage{soul}

\graphicspath{{/Users/p_salati/Dropbox/CRAC_3_positrons/CRAC_3_draft/figures/},{figures/},{./}}
\usepackage{txfonts}
\usepackage{natbib,twoopt}
\usepackage[breaklinks=true]{hyperref} 
\bibpunct{\textcolor{black}{(}}{)}{;}{a}{}{,}             
\makeatletter

\newcommandtwoopt{\citeads}[3][][]{\href{http://adsabs.harvard.edu/abs/#3}
  {\def\hyper@linkstart##1##2{\color{black}}
  \let\hyper@linkend\@empty\citealp[#1][#2]{#3}}}

\newcommandtwoopt{\citepads}[3][][]{\href{http://adsabs.harvard.edu/abs/#3}
  {\def\hyper@linkstart##1##2{\color{black}}
  \let\hyper@linkend\@empty\citep[#1][#2]{#3}}}

\newcommandtwoopt{\citetads}[3][][]{\href{http://adsabs.harvard.edu/abs/#3}
  {\def\hyper@linkstart##1##2{\color{black}}
  \let\hyper@linkend\@empty\citet[#1][#2]{#3}}}

\newcommandtwoopt{\citeyearads}[3][][]
  {\href{http://adsabs.harvard.edu/abs/#3}
  {\def\hyper@linkstart##1##2{\color{black}}
  \let\hyper@linkend\@empty\citeyear[#1][#2]{#3}}}

\makeatother

\usepackage{hyperref}
\usepackage{hypcap}
\definecolor{cyan}{cmyk}{1.,0.,0.,0.2}
\definecolor{vert}{cmyk}{0.5,0.,0.5,0.5}
\definecolor{magenta}{cmyk}{0.,1.,0.,0.1}
\definecolor{verdatre}{cmyk}{0.5,0.,0.5,0.5}
\definecolor{vert_clair}{cmyk}{0.5,0.,0.5,0.2}
\definecolor{yellow}{cmyk}{0.,0.,1.,0.0}
\definecolor{yellow_1}{cmyk}{0.,0.,0.5,0.0}
\definecolor{rouge}{cmyk}{0.,0.4,0.6,0.0}
\definecolor{orange}{cmyk}{0.,0.5,0.5,0.05}
\definecolor{violet}{rgb}{0.5,0.,0.5}
\definecolor{darwin_box}{rgb}{0.988,0.878,0.77}
\definecolor{darwin_text}{rgb}{0.1,0.07,0.02}
\usepackage{multicol}
\definecolor{rouge1}{cmyk}{0,1,1,0.4}
\definecolor{rouge2}{cmyk}{0,1,1,0.55}
\definecolor{rouge3}{cmyk}{0,1,1,0.2}
\definecolor{bleu1}{cmyk}{1,1,0,0.3}
\definecolor{bleu2}{cmyk}{1,1,0,0.6}
\definecolor{bleu3}{cmyk}{1,1,0,0.1}
\definecolor{vert1}{cmyk}{0.92,0,0.59,0.25}
\definecolor{vert2}{cmyk}{0.92,0,0.59,0.15}
\definecolor{vert3}{cmyk}{0.92,0,0.59,0.5}
\newcommand{\beq}{\begin{equation}}
\newcommand{\eeq}{\end{equation}}
\newcommand{\bea}{\begin{eqnarray}}
\newcommand{\ena}{\end{eqnarray}}

\newcommand{\ie}{i.e.}

\newcommand{\bal}{\begin{aligned}}
\newcommand{\eal}{\end{aligned}}

\newcommand{\Vc}{\ensuremath{V_c}\xspace}
\newcommand{\vecVc}{\ensuremath{\vec{V}_c}\xspace}
\newcommand{\Va}{\ensuremath{V_a}\xspace}

\def\HEAT{{\sc Heat}}
\def\PAMELA{{\sc Pamela}}
\def\AMS{{\sc Ams-02}\xspace}
\def\HEA{{\sc Hea03}\xspace}

\def\CREAM{{\sc Cream}}

\def\MIN{{\sc Min}}
\def\MED{{\sc Med}}
\def\MAX{{\sc Max}}
\def\beff{b^{\rm eff}_{\rm halo}}
\newcommand\CellTop{\rule{0pt}{2.2ex}}

\begin{document}

\title{The pinching method for Galactic cosmic ray positrons: \\ implications in the light of precision measurements}

\author{
M.~Boudaud \inst{1,4}
\and E.~F.~Bueno\inst{3}
\and S.~Caroff \inst{2}
\and Y.~Genolini \inst{1}
\and V.~Poulin \inst{1,5}
\and\\
V.~Poireau \inst{2}
\and A.~Putze \inst{1,2}
\and S.~Rosier \inst{2}
\and P.~Salati \inst{1}
\and M.~Vecchi \inst{3}
\fnmsep\thanks{The authors are members of the Cosmic Ray Alpine Collaboration.}
\fnmsep\thanks{Contact authors are mathieu.boudaud@lapth.cnrs.fr, sami.caroff@lapp.in2p3.fr, yoann.genolini@lapth.cnrs.fr, and vivian.poulin@lapth.cnrs.fr}
}

\institute{
Laboratoire d'Annecy-le-Vieux de Physique th\'eorique (LAPTh), CNRS and Universit\'e Savoie Mont Blanc, 9 Chemin de Bellevue, B.P.110 Annecy-le-Vieux, F-74941, France\\
%
\and
Laboratoire d'Annecy-le-Vieux de Physique des Particules (LAPP), CNRS/IN2P3 and Universit\'e Savoie Mont Blanc, 9 Chemin de Bellevue, B.P.110 Annecy-le-Vieux, F-74941, France\\
%
\and
Instituto de F\'isica de S\~ao Carlos (IFSC), Universidade de S\~ao Paulo, CP 369, 13560-970,  S{\~a}o Carlos, SP,~Brazil\\
%
\and
Laboratoire de Physique Th\'eorique et Hautes \'Energies (LPTHE), CNRS and Universit\'e Pierre et Marie Curie, Bo\^{\i}te 126, T13-14 4\`eme \'etage,
4 place Jussieu 75252 Paris cedex 05, France\\
%
\and
Institute for Theoretical Particle Physics and Cosmology (TTK), RWTH Aachen University, D-52056 Aachen, Germany\\
%
}

\date{Received; accepted\\
Preprint numbers : LAPTH-TBD/16}


\abstract
%
%
{Two years ago, the \AMS\ collaboration released the most precise measurement of the cosmic ray positron flux. In the conventional approach, in which positrons are considered as purely secondary particles, the theoretical predictions fall way below the data above 10 GeV. One suggested explanation for this anomaly is the annihilation of dark matter particles, the so-called WIMPs, into standard model particles. Most analyses have focused on the high-energy part of the positron spectrum, where the anomaly lies, disregarding the complicated GeV low-energy region where Galactic cosmic ray transport is more difficult to model and solar modulation comes into play.}
%
%
{Given the high quality of the latest measurements by  \AMS, it is now possible to systematically re-examine the positron anomaly over the entire energy range, this time taking into account transport processes so far neglected, such as Galactic convection or diffusive re-acceleration. These might impact somewhat on the high-energy positron flux so that a complete and systematic estimate of the secondary component must be performed and compared to the \AMS\ measurements. The flux yielded by WIMPs also needs to be re-calculated more accurately to explore how dark matter might source the positron excess.
}
%
%
{We devise a new semi-analytical method to take into account transport processes so far neglected, but important below a few GeV. It is essentially based on the pinching of inverse Compton and synchrotron energy losses from the magnetic halo, where they take place, inside the Galactic disc. The corresponding energy loss rate is artificially enhanced by the so-called pinching factor which needs to be calculated at each energy. We have checked that this approach reproduces the results of the Green function method at the per mille level. This new tool is fast and allows to carry out extensive scans over the cosmic ray propagation parameters.}
%
%
{We derive the positron flux from sub-GeV to TeV energies for both gas spallation and dark matter annihilation. We carry out a scan over the cosmic ray propagation parameters which we strongly constrain by requiring that the secondary component does not overshoot the \AMS\ measurements. We find that only models with large diffusion coefficients are selected by this test.
We then add to the secondary component the positron flux yielded by dark matter annihilation. We carry out a scan over WIMP mass to fit the annihilation cross section and branching ratios, successively exploring the cases of a typical beyond-the-standard-model WIMP and an annihilation through light mediators. In the former case, the best fit yields a $p$-value of 0.4\% for a WIMP mass of 264~GeV, a value that does not allow to reproduce the highest energy data points. If we require the mass to be larger than 500~GeV, the best-fit $\chi^{2}$ per degree of freedom always exceeds a value of 3. The case of light mediators is even worse, with a best-fit $\chi^{2}$ per  degree of freedom always larger than 15.}
%
%
{We explicitly show that the cosmic ray positron flux is a powerful and independent probe of Galactic cosmic ray propagation. It should be used as a complementary observable to other tracers such as the boron-to-carbon ratio. This analysis shows also that the pure dark matter interpretation of the positron excess is strongly disfavored. This conclusion is based solely on the positron data, and no other observation, such as the antiproton flux or the CMB anisotropies, needs to be invoked.}

\maketitle

\newpage

\section*{}
\textcolor{white}{blabla}


\section{Introduction}

The cosmic ray (CR) positron flux has been measured with unprecedented accuracy by the \AMS\ collaboration (\cite{2014PhRvL.113l1102A}). This observation is of paramount importance in several respects.
To start with, it provides an insight into the mechanisms that create positrons inside the Milky Way. For a long time, CR positrons have been thought to be exclusively secondary species originating from the spallation of the gas lying in the Galactic disc. The first experimental hints for a deviation from the conventional scenario came from the data collected by the High-Energy Antimatter Telescope (\HEAT) collaboration\citepads{1997ApJ...482L.191B,2001ApJ...559..296D,2004PhRvL..93x1102B}, but the existence of a positron anomaly has been firmly established by~\citetads{2009Natur.458..607A} who reported an excess in the positron fraction measured up to 100~GeV by the Payload for Antimatter Matter Exploration and Light-nuclei Astrophysics (\PAMELA) satellite. Recently, the Alpha Magnetic Spectrometer (\AMS) has initiated a new era of precision measurements with the release of high-quality data, in particular on the positron fraction\citepads{2013PhRvL.110n1102A,2014PhRvL.113l1101A} and positron flux\citepads{2014PhRvL.113l1102A} up to 500~GeV. The \AMS\ results definitely confirm that, in addition to the secondary component, a new ingredient is at play in the cosmic positron radiation.

\vskip 0.1cm
These measurements have an obvious connection with the long standing problem of the astronomical dark matter (DM). The nature of this enigmatic component, which contributes a fraction of $\sim$27\% to the energy budget of the Universe, is still unresolved. The most commonly accepted hypothesis is based on a weakly interacting massive particle, dubbed Weakly Interacting Massive Particle (WIMP), whose existence is predicted in most extensions of the high-energy physics standard model. A distinctive feature of these species is to be produced in the early Universe, through self-annihilation, with a relic abundance in close agreement with the cosmological observations. In this approach, WIMPs pervade the Galactic halo wherein they still pair annihilate today, yielding positrons among other standard model particles. The positron anomaly has triggered a feverish activity insofar as it could be interpreted as the imprint left by DM species on the CR positron spectrum. Many investigations explored whether or not WIMPs might be the source of that anomaly. We refer the reader to the analysis by\citepads{2014JCAP...04..006D,2016JCAP...05..031D,2015PhRvD..91f3508L,2015A&A...575A..67B} and references therein. The vast majority of these studies are focused on the high-energy part of the positron spectrum, above 10 GeV. Below this energy, solar modulation comes into play and complicates the interpretation of the data. Moreover, Galactic convection, diffusive re-acceleration, and positron annihilation on interstellar gas must be taken into account in addition to space diffusion. Finally, energy losses, which play a key role in the propagation of positrons, are mostly concentrated inside the Galactic disc whereas they extend all over the magnetic halo at high energy.

\vskip 0.1cm
Besides the complexity of modelling CR transport below a few GeV, there is also the implicit but wide spread assumption that DM should essentially show up at high energies. The corollary of that standpoint is that some yet-to-be-determined astrophysical sources should be operating at low energies so as to produce the positron flux at the observed level. Pulsars\citepads{2009JCAP...01..025H,2012CEJPh..10....1P,2013ApJ...772...18L} or spallations inside the supernova driven shock waves\citepads{2009PhRvL.103e1104B,2014PhRvD..90f1301M} are two examples of such possibilities. But if additional processes need to be invoked to explain the low-energy part of the positron data, nothing precludes them from coming into play also at higher energies. There is nothing special taking place at a scale of a few GeV, and processes known to be active at low energy are expected to contribute all over the positron spectrum.
Of course, looking for a DM solution of the positron anomaly by fitting the WIMP cross section and mass to the high-energy part of the data is tantalizing. This has actually been the subject of numerous studies since the discovery of the positron excess by {\PAMELA}. But these analyses are based on the prejudice that low-energy positron data are not relevant to DM, an unwarranted assumption that might introduce biases.
For instance, should some WIMP parameters provide a good fit, one might be left with the illusion that the positron excess is a smoking gun signature of the presence of DM species inside the Galaxy. One should instead conclude that, even though the data actually support the WIMP hypothesis, they cannot be considered as a compelling evidence for particle DM. For this, DM annihilation and gas spallation by CR nuclei must be proved to be the only sources of high-energy positrons while, at the same time, other production mechanisms are assumed to operate at low energy.
Another flaw in this approach is the risk that a WIMP model that fits well the positron excess above a few GeV could simultaneously be a poor match to the low-energy data, overshooting them for instance.
Establishing the DM hypothesis requires then to derive the positron flux over the entire accessible energy range, and not just on its high-energy part.

\vskip 0.1cm
These considerations have led us to reinvestigate the problem of the positron anomaly over the entire energy range covered by the \AMS\ data. In order to test the DM hypothesis, we have computed the interstellar positron flux yielded at the Earth by (i) the spallation of interstellar gas by CR protons and helium nuclei and (ii) WIMP annihilation. To do so, we have conceived a new semi-analytical method to cope with transport processes so far neglected but important below a few GeV. This new tool allows also to carry out extensive scans over the CR propagation parameters. The main point of novelty consists in the analytic treatment of the energy losses suffered by cosmic ray positrons in the magnetic halo: the corresponding effect being artificially enhanced by the so-called pinching factor, that shifts the energy losses from the magnetic halo, where they actually take place, inside the Galactic disc. An essential benefit is a faster runtime in comparison to a fully numerical approach.

\vskip 0.1cm
We have also improved the calculation of the background of secondary positrons by using the new measurements of the CR proton and helium fluxes by the \AMS\ collaboration\citepads{2015PhRvL.114q1103A,2015PhRvL.115u1101A}. There is a hardening above $\sim$300~GeV that leads to a slight increase of the positron yield from gas spallation.
We have overcome the difficulty arising from solar modulation by using the value of the Fisk potential inferred by\citetads{2016A&A...591A..94G} from their analysis of the variations of the top-of-the-atmosphere (TOA) proton flux over the recent past.

\vskip 0.1cm
Equipped with the pinching method, we have improved upon the analysis by\citetads{2014PhRvD..90h1301L} by carrying out a scan over the CR propagation models found by\citetads{2001ApJ...555..585M} to be compatible with the B/C ratio, and by deriving for each of them the positron flux yielded by gas spallation.
We have finally investigated the DM solution to the positron anomaly by calculating, for each of the surviving CR models, the yield from an annihilating WIMP to which we have added the secondary positron background. The positron flux is derived over the same energy range as for the \AMS\ data. We have performed a scan over WIMP mass and explored the possibility of mixed annihilation channels. At fixed WIMP mass, we have derived the best-fit values of the annihilation cross section and branching ratios. We have considered DM particles annihilating either directly in standard model particles or through light mediators.

\vskip 0.1cm
The article is organized according the points sketched above.
The pinching method, which is paramount to this investigation, is detailed in Sect.~\ref{sec:pinching}.
We apply this new tool in Sect.~\ref{sec:implication} to investigate the implications on the positron flux of CR transport processes so far neglected at high energies. The astrophysical background of secondary positrons is discussed in Sect.~\ref{sec:astrobkg} while the positron flux yielded by DM species is presented in Sect.~\ref{prim:DM_species}.
We then constrain in Sect.~\ref{sec:prop_param} the CR propagation parameters, requiring that they do not lead to a flux of secondary positrons in excess of the measurements. The scan procedure is exposed and results into a sharp selection of the CR models.
The DM interpretation of the \AMS\ data is presented in Sect.~\ref{sec:DM_interpretation}.
In Sect.~\ref{sec:robustness} we investigate the robustness of our results and explore a few sources of uncertainties.
We finally conclude in Sect.~\ref{sec:conclusion}.



\section{Propagation of Galactic cosmic ray positrons with the pinching method}
\label{sec:pinching}

In this section, we recall the basics of the propagation of CRs in the Galaxy. We first present the transport equation and its semi-analytical resolution.We then introduce a new method, called hereafter the \textit{pinching method}, to solve semi-analytically the transport equation for electrons and positrons when {\em all} propagation effects are simultaneously taken into account.


\subsection{The transport equation of CRs}

During their journey across the Galaxy, CRs are affected by many processes as a result of their interactions with the Galactic magnetic field (GMF) and the interstellar medium (ISM).
Despite the strength of the magnetic turbulence, Fick's law still holds \citepads{2002PhRvD..65b3002C}.
Hence, the scattering of CRs on the GMF can be described by a random walk and modelled by a diffusion process in space. In this work we choose an homogeneous and isotropic diffusion coefficient $K(E) = \beta K_0 (\mathcal{R}/1 \, \rm{GV})$ where $\beta$ is the velocity of the particle and $\mathcal{R}$ the rigidity related to its momentum $p$ and its charge $q$ by $\mathcal{R} = p/q$.
On top of this, the diffusion centers 
move with the Alfv\`{e}n waves velocity \Va. Thus, CRs undergo a diffusive reacceleration (DR) thanks to the second-order Fermi mechanism. This process can be modelled by a diffusion in energy space with coefficient $D(E) = (2/9) \Va^2 E^2 \beta^4 / K(E).$
Moreover, CRs can interact with the ISM, leading to: i) energy losses from Coulomb interaction and ionisation, with respective rates $b_{\rm coul}$ and $b_{\rm ioni}$; ii) their destruction at a rate $\Gamma$.
In addition, electrons and positrons (loosely dubbed electrons hereafter except when explicitly mentioned) lose energy by bremsstrahlung, synchrotron emission as well as inverse Compton (IC) scattering when they interact with the interstellar radiation field (ISRF), at respective rates $b_{\rm brem}$, $b_{\rm sync}$, and  $b_{\rm IC}$.
Following the procedure described in\citetads{2010A&A...524A..51D}, we consider IC scattering in the relativistic regime and make use of the mean value of the GMF $\langle B \rangle = 1 \, \rm{\mu G}$\citepads{2001RvMP...73.1031F}.
Finally, CRs are blown by the Galactic wind (GW) produced by supernova remnant explosions in the Galactic disc. We assume the GW to be homogeneous and perpendicular to the Galactic disc, with velocity $\vecVc = \text{sign}(z) \Vc \, \vec{e}_z$. This process leads to the adiabatic cooling of CRs, which enters as an additional term in the energy loss rate $b_{\rm adia}$.
The total energy loss rate $b(E) \equiv dE/dt$ is then simply the sum of all the loss processes (their explicit expression can be found in\citetads{1998ApJ...509..212S} and\citetads{1998ApJ...493..694M}).

\vskip 0.1cm
Following the work of\citetads{2001ApJ...555..585M} (and reference therein), we assume the Galaxy to be axisymmetric and describe it by the two-zone model. The first zone, within which ISM is homogeneously distributed, represents the Galactic disc of half-height $h = 100 \, \rm{pc}$. Atomic densities are taken to be $n_{\rm H} = 0.9 \, \rm{cm^{-3}}$ and $n_{\rm He} = 0.1 \, \rm{cm^{-3}}$. It is embedded inside a much larger second zone, namely the magnetic halo (MH), of half-height $L$ lying between 1 and 15 kpc. Both zones share the radius $R = 20 \, \rm{kpc}$.
In practice, we assume the space diffusion, as well as the energy losses from synchroton emission and IC scattering, to lie in the whole magnetic halo. On the other hand, DR, bremsstrahlung, Coulomb interaction, ionisation, and destruction take place only in the Galactic disc where the matter of the ISM is concentrated\citepads{1997A&A...321..434P}. Hence, we split the energy losses $b(E,z)$ into a disc component $b_{\rm disc} \equiv b_{\rm coul} + b_{\rm brem} + b_{\rm ioni} + b_{\rm adia}$ that includes the mechanisms that take place only in the Galactic disc, and a halo component $b_{\rm halo} \equiv b_{\rm IC} + b_{\rm sync}$ considering those that take place in the whole magnetic halo (including the disc). We impose a vanishing density of CRs outside the MH of the Galaxy.

Under a steady state and thin disc approximation, the density of CRs per unit of space and energy $\psi \equiv dN/d^3 \! x  dE$ obeys the transport equation
\beq
\label{eq:full_transport}
\bal
&\vec{\nabla} \cdot \left[ \vecVc \, \psi(E, r, z) - K(E) \, \vec{\nabla} \psi(E, r, z) \right]   +  \\ &\partial_E \left[ b(E,z) \, \psi(E, r, z) -  \, 2h \, \delta(z) \, D(E) \, \partial_{E}\psi(E, r, z) \right] + \\   &2h \, \delta(z) \, \Gamma \psi = Q(E, r, z),
\eal
\eeq
where $Q$ represents the injection rate of CRs in the Galaxy.

\vskip 0.1cm
CR nuclei lose energy only in the Galactic disc (\ie\ $b_{\rm halo}^{\rm nuc} = 0$). In this case, the transport equation~(\ref{eq:full_transport}) can be solved via the semi-analytical scheme introduced in\citetads{2001ApJ...555..585M}.
More precisely, the CR density $\psi$ is expanded on the basis of the first-order Bessel functions $J_0$ such that
\beq
\psi(E,r,z) = \sum_{i=1}^{\infty} J_0 \left( \alpha_i \frac{r}{R} \right) \, P_i(E,z), \\
\eeq
where $\alpha_i$ are the zero of the Bessel function $J_0$. The transport equation~(\ref{eq:full_transport}) becomes consequently
\beq
\bal
\label{eq:full_transport_bessel}
&\partial_z[ \Vc (z) \, P_i] - K(E) \,  \partial^{2}_{z} P_i + K(E) \left( \frac{\alpha_i}{R} \right)^2 P_i + \\ &2h \, \delta(z) \, \partial_E \left[ b(E) \, P_i - \, D(E) \, \partial_{E} P_i \right] + 2h \, \delta(z) \, \Gamma P_i = Q_i(E, z),
\eal
\eeq
where $Q_i(E, z)$ are the Bessel transform coefficients of the source term $Q(E,  r, z)$.
Eq.~(\ref{eq:full_transport_bessel}) is then reduced to a second-order ordinary differential equation for the function $P_i(E, z=0)$ with respect to the energy $E$, and can be solved numerically using a Cranck-Nicholson algorithm.
Finally, the CR flux at the Earth is given by $\Phi(E, \odot) = v/4\pi  \,  \psi (E, \odot)$ where $r_{\odot} = 8.5 \; \rm kpc$. For more details on the resolution method, we refer the reader to\citetads{2001ApJ...555..585M}. In this way, previous authors used the semi-analytical method to determine 1,623 sets of propagation parameters constrained by the boron over carbon ratio B/C measurements. This enabled them to derive in\citetads{2004PhRvD..69f3501D} the benchmark \MIN, \MED, and \MAX\ propagation models presented in Table~\ref{tab:Donato2005}.
%
\begin{table*}[h!]
\begin{center}
\caption{\small Benchmark \MIN, \MED, and \MAX\ sets of propagation parameters introduced in\citetads{2004PhRvD..69f3501D}.}
\begin{tabular}{cccccc}
\hline
\hline
\CellTop
Case		& $\delta$			& $K_0$~[kpc$^2$/Myr]		& $L$~[kpc]		& \Vc~[km/s]	&  \Va~[km/s]\\
\hline
\CellTop
\MIN  		& 0.85			& 0.0016					& 1				& 13.5			&  22.4 \\
\MED  		& 0.70			& 0.0112					& 4 	 		& 12	 		& 52.9 \\
\MAX  		& 0.46			& 0.0765					& 15 			& 5	 			& 117.6\\
\hline
\end{tabular}
\label{tab:Donato2005}
\end{center}
\end{table*}
%

\vskip 0.1cm
In the case of electrons, the semi-analytical resolution of the transport equation, {\em as it is}, is not possible. Indeed, the difficulty comes from the fact that electrons lose energy in the Galactic disc as well as in the whole magnetic halo. In the thin disc approximation, the energy loss rate can be written $b(E,z) = 2h \, \delta(z) \, b_{\rm disc}(E) + b_{\rm halo}(E)$, but the presence of the term $b_{\rm halo}$ prevents direct semi-analytical resolution of Eq.~(\ref{eq:full_transport_bessel}). Therefore, numerical codes have been adopted to predict the flux of electrons at the Earth.
An alternative way, often used in literature, is to focus only on high-energy electrons ($E$ > few GeV). In this case, as shown in \citetads{2009A&A...501..821D}, the dominant propagation processes are the space diffusion and the halo energy losses ($b_{\rm sync}$ and $b_{\rm IC}$). The high-energy approximation consists thus in neglecting the DR, the convection, the disc energy losses $b_{\rm disc}$, and the destruction of CRs. Hence, the high-energy transport equation can be written
\beq
\label{eq:HE_transport}
 - K(E) \, \Delta \psi  + \partial_E \left[ b(E) \, \psi \right] = Q(E, r, z),
\eeq
where $b = b_{\rm halo}$.
Eq.~(\ref{eq:HE_transport}) can be solved analytically using the pseudo-time method introduced by\citetads{1999PhRvD..59b3511B} and its solution can be expressed in term of Bessel functions\citepads{2008PhRvD..77f3527D}, where the Bessel coefficients evaluated at $z=0$ are given by
\beq
\label{eq:Pi_halo}
P_i(E, 0) = \frac{-1}{b(E)} \int \limits_E^{+\infty} \!\! dE_S \; B_i(E, E_S),
\eeq
where
\beq
B_i(E, E_S) = \sum \limits_{n=2m+1}^{+\infty} Q_{i,n}(E_S) \exp\left[ - C_{i,n} \lambda_D^2 \right].
\eeq
The function $Q_{i,n}$ is the Fourrier transform of $Q_i(E, z)$ defined as
\beq
Q_{i,n}(E) 	= \frac{1}{L} \int \limits_{-L}^{L} \!\! dz \; \varphi_n(z) \, Q_i(E, z),
\eeq
where $\varphi_n(z) = \cos(nk_0 \, z)$ with $k_0 = \pi /2L$. The coefficient $C_{i,n}$ is defined as
\beq
C_{i,n} = \frac{1}{4} \left[ \left( \frac{\alpha_i}{R} \right)^2 + (n k_0)^2 \right].
\eeq
Finally, the diffusion length $\lambda_D$ is related to the space diffusion coefficient $K$ and the energy loss rate $b$ by the expression
\beq
\lambda^2_D(E, E_S) = 4 \int \limits_{E_S}^E \!\! dE' \; \frac{K(E')}{b(E')}.
\eeq
Note that the density $\psi$ at the Earth can be written as
\beq
\psi(E, \odot) = \frac{-1}{b(E)} \int \limits_{E}^{+\infty}  \!\! dE_{S} \; \mathcal{I}(\lambda_D) \, Q(E_S, \odot),
\eeq
where the halo integral $\mathcal{I}$ is defined as
\beq
\mathcal{I}(\lambda_D) =  \sum \limits_{i=1}^{+\infty} J_0\left( \alpha_i\frac{ r}{R} \right) \frac{B_{i}(E, E_S)}{Q(E_S, \odot)}.
\eeq

The flux at the Earth can then be computed for secondary electrons from proton and helium spallation\citepads{2009A&A...501..821D,2015A&A...575A..67B}, as well as for primary electrons from the ones produced by DM particle annihilations\citepads{2008PhRvD..77f3527D,2015A&A...575A..67B} and astrophysical objects like pulsars\citepads{2015A&A...575A..67B}. One can then perform comparisons with data, which have led to the discovery of a high-energy positron excess requiring the presence of a dominant primary component above approximately 10 GeV.
The high-energy approximation is often used in the literature to derive conclusions for energies above that value. However, it is not obvious that the low-energy propagation effects (DR, convection, and energy losses in the Galactic disc) can be safely neglected, especially in the era of the \AMS high-accuracy measurements.
Furthermore, due to high statistics, the region below 10 GeV is affected by the lowest experimental uncertainties and could thus provide the strongest constraints. These considerations led us to develop a new theoretical solution for the propagation of electrons over the energy range covered by \AMS.
This method dubbed \textit{pinching method} is described in the following section.


\subsection{The pinching method}

At first sight, it seems that the semi-analytical method cannot be used to solve Eq.~(\ref{eq:full_transport}) when energy losses take place simultaneously in the MH and in the Galactic disc.
The trick to overcome this issue is to impose the halo energy losses to take place, in an effective way, only in the Galactic disc. In other words, it consists in replacing the term $b_{\rm halo}$ in the transport equation~(\ref{eq:full_transport}) with an effective term $2h \, \delta(z) \, \beff$ while keeping the same solution $\psi$. By doing so, it will be possible to rewrite Eq.~(\ref{eq:full_transport}) in the form of Eq.~(\ref{eq:full_transport_bessel}) and to apply the Crank-Nicholson algorithm to solve it.
This procedure consists thus in {\em pinching} the halo energy losses inside the disc, hence the name \textit{pinching method}.

\vskip 0.1cm
The function $\beff$ depends on all propagation effects electrons undergo. Nevertheless, from few GeV to 1 TeV, halo energy losses and space diffusion are the dominant propagation processes\citepads{2009A&A...501..821D}. Hence, at first order, we can reasonably neglect other processes and determine $\beff$ using the high-energy approximation, \ie, Eq.~(\ref{eq:HE_transport}).
This approximation may not be completely valid for energies below a few GeV where other effects come into play and are expected to affect the calculation of $\beff$. But the more dominant these processes are, the less important halo energy losses turn to be, so that the precise value of the pinching factor does not matter at low energies.

\vskip 0.1cm
Let us start with the pedagogical case of a monochromatic source of electrons $Q(E, r, z) = \delta(E - E_S) \, Q(r, z)$.
In order to determine $\beff$, we compute first the exact high-energy solution $\psi^h$ using the pseudo-time method described above. The index $h$ means that $\psi^h$ is solution of Eq.~(\ref{eq:HE_transport}) where IC and synchrotron energy losses are distributed in the whole MH.  In that case, the electron density $\psi^h$ at $z = 0$ is given by
\beq
\psi^h(E, r, 0) = \sum \limits_{i=1}^{+\infty} J_0 \left( \alpha_i \frac{r}{R} \right) P_{i}^{\, h}(E, 0),
\eeq
where $P_{i}^{\, h}(E, 0)$ is given by the expression~(\ref{eq:Pi_halo}).\\
In a second step, we introduce $\psi^d$, solution of the high-energy equation
\beq
\label{eq:HE_transport_disc}
 - K(E) \, \Delta \psi^d  + 2h \, \delta(z) \, \partial_E \left[ \beff \, \psi^d \right] = Q(E, r, z),
\eeq
where IC and synchrotron energy losses are confined to the disc. The condition $\psi^h(E, r, 0) = \psi^d(E, r, 0)$  enables then to determine the function $\beff$ such that
\beq
\beff(E, E_S, r) = \xi(E, E_S, r) \, b_{\rm halo}(E),
\eeq
where we introduced the pinching factor $\xi(E, E_S, r)$, given by the expression
\beq
\xi(E, E_S,  r) = \frac{1}{\psi^h(E, r, 0)}  \sum_{i=1}^{+\infty} J_0 \left( \alpha_i \frac{r}{R} \right) \,  \xi_i(E, E_S) \, P_i(E,0),
\eeq
with
\beq
\xi_i(E, E_S) = \frac{1}{B_i(E, E_S)} \left[ J_i(E_S) + 4 \, k_i^2 \int \limits_{E}^{E_S} \!\!\! dE' \frac{K(E')}{b_{\rm halo}(E')} \; B_i(E', E_S) \right].
\eeq
The coefficient $J_i$ and $k_i$ are given by
\beq
k_i^2 = \frac{S_i}{8h} \coth \left( \frac{S_i \, L }{2} \right).
\eeq
and
\beq
J_i(E_S) = \frac{1}{h} \int \limits_0^{L}  \!\! dz_S \; \mathcal{F}_i(z_S) \, Q_i(E_S, z_S),
\eeq
where $S_i \equiv 2 \alpha_i /R$ and
\beq
\mathcal{F}_i(z) = {\sinh \left[ \dfrac{S_i}{2}(L - z) \right]}/{\sinh \left[ \dfrac{S_i \, L}{2} \right]}.
\eeq
Once the effective term $\beff$ has been computed, it is possible to switch on low-energy effects and to solve Eq.~(\ref{eq:full_transport_bessel}) with {\em all} propagation processes using the usual Crank-Nicholson algorithm.

\vskip 0.1cm
In practice, the electron source term is not a Dirac function but follows a continuum distribution in energy, which depends on the actual source considered (e.g. spallation in the disc, DM, pulsars). We therefore have to compute the pinching coefficients $\xi_i(E, E_S)$ for each electron energy at source $E_S$, which requires a very long computational time. However, an alternative way consists in averaging the quantity $\xi_i(E, E_S)$ over electron energies at source $E_S$. We show in Sec.\ref{sec:testing_pinching} that the effect arising from this simplification is kept below 0.2\% over the whole energy range. We describe in the following how to perform such averaging.

\begin{figure*}[h!]
\begin{center}
\includegraphics[width=0.5\textwidth]{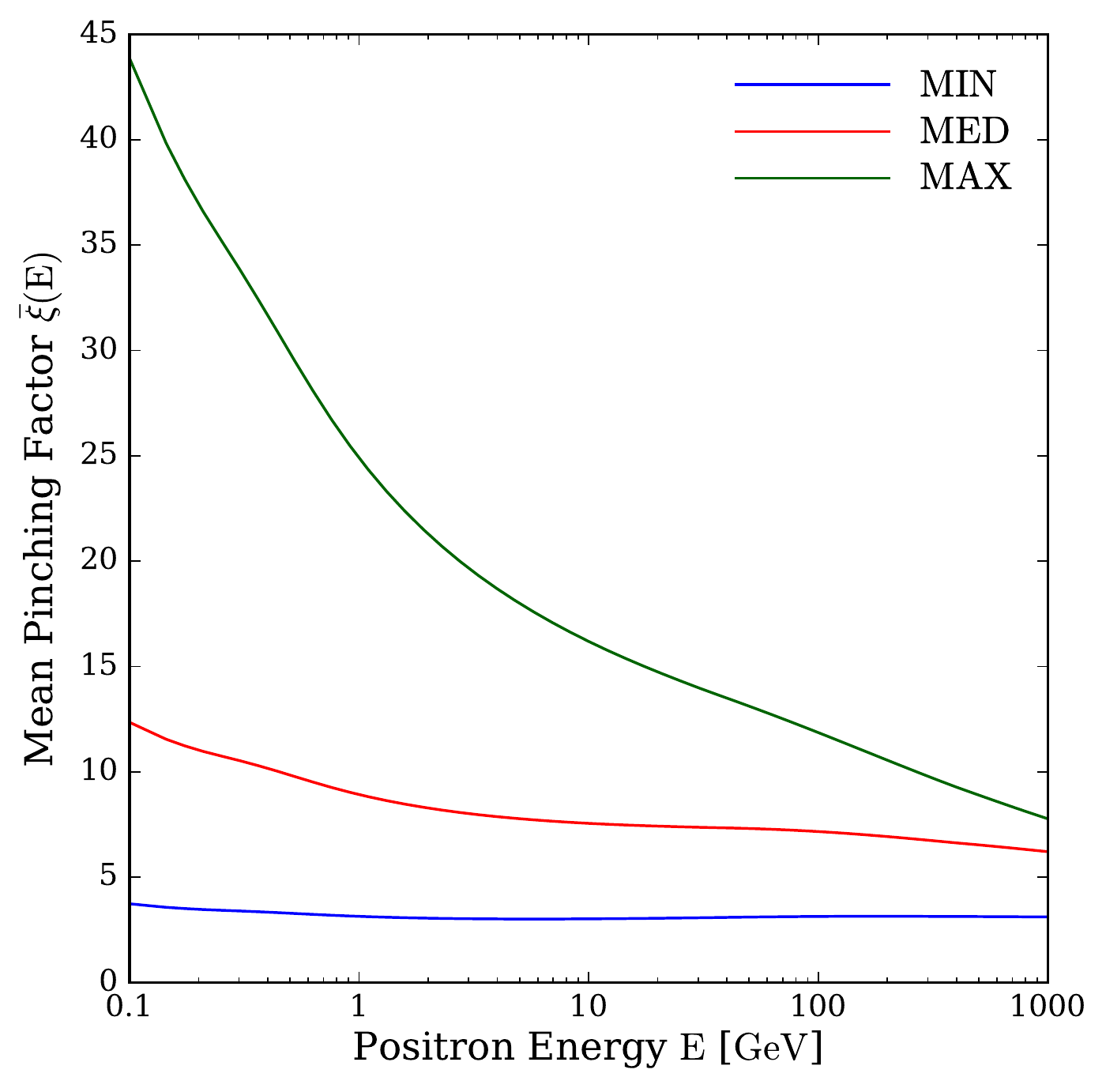}
\caption{The mean pinching factor of secondary positrons computed for the \MIN\ (blue), \MED\ (red), and \MAX\ (green) models as a function of the positron energy.}
\label{fig:sec_pinching_MIN_MED_MAX}
\end{center}
\end{figure*}

\vskip 0.1cm
Let $\mathcal{P}_i(E, E_S) = p_i(E, E_S) \, dE_S$ be the probability that an electron, injected with energy in the range $[E_S,E_S + dE_S]$ and measured at the Earth with an energy $E$,  contributes to the $i^{\rm th}$ Bessel order of the Bessel transform $P^{\, h}_i(E,0)$. The associated probability density $p_i$ is then given by
\beq
p_i(E, E_S) = \frac{B_i(E, E_S) }{\int \limits_{E}^{+\infty}  \!\! dE_S \, B_i(E, E_S)}.
\eeq
Therefore, the mean value of the pinching coefficients $\xi_i(E, E_S)$ is given by the expression
\beq
\label{eq:xi_moyen_general}
\bar{\xi}_i(E) = \frac{\int \limits_E^{+\infty}  \!\! dE_S \left[ J_i(E_S) + 4k_i^2 \int \limits_E^{E_S} \! \! dE' \, \dfrac{K(E')}{b(E')} \, B_i(E', E_S) \right]  }{ \int \limits_E^{+\infty} dE_S \, B_i(E, E_S) },
\eeq
and
\beq
\label{eq:xi_bar_r}
\bar{\xi}(E, r) = \frac{1}{\psi^h(E, r, 0)}  \sum_{i=1}^{+\infty} J_0(\alpha_i \frac{r}{R}) \,  \bar{\xi}_i(E) \, P_i(E,0).
\eeq
The mean pinching factor $\bar{\xi}(E)$ of secondary positrons is represented in Fig.\ref{fig:sec_pinching_MIN_MED_MAX} for the \MIN, \MED\ and \MAX\ sets of propagation parameters. As it is expected the pinching factor is larger in the case of \MAX, that corresponds to the larger value of $K_0$ and $L$, where the 
effect of the pinching must be more important.
%
\vskip 0.1cm


\subsection{Testing the pinching method}
\label{sec:testing_pinching}

We wish to assess the theoretical uncertainty of the pinching method used to compute the positron flux. We focus our study on the energy range probed by \AMS i.e. the rough interval [100 MeV, 1 TeV].
To this aim, we compare the analytical solution of Eq.~(\ref{eq:HE_transport}) to the semi-analytical solution arising from the pinching method Eq.~(\ref{eq:HE_transport_disc}).
Thus, we switch off the low-energy processes (DR, disc energy losses, convection, and destruction) and consider only halo energy losses and space diffusion processes (high-energy approximation).

\begin{figure*}[h!]
\begin{center}
\includegraphics[width=0.47\textwidth]{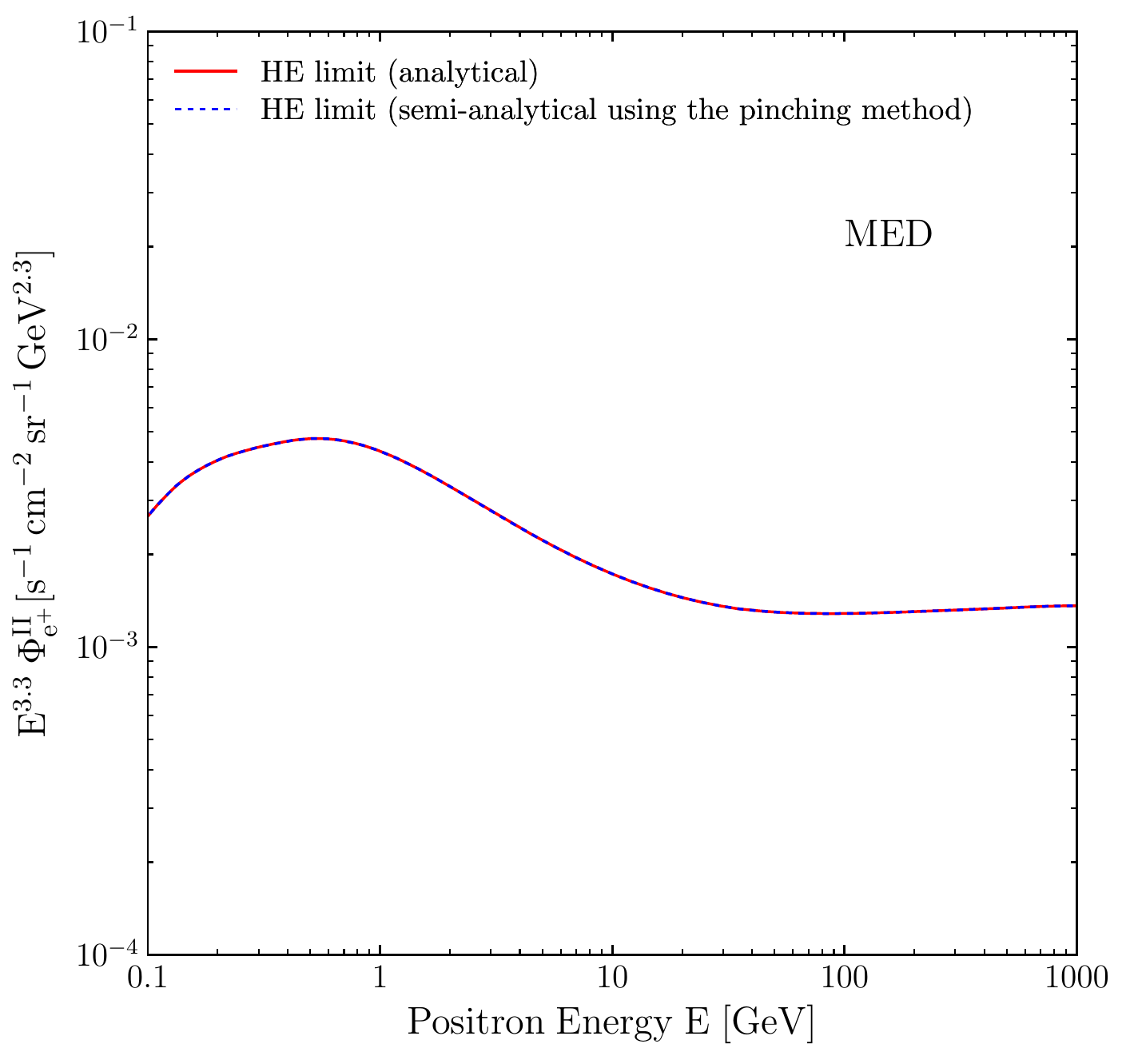}
\hspace{1cm}
\includegraphics[width=0.46\textwidth]{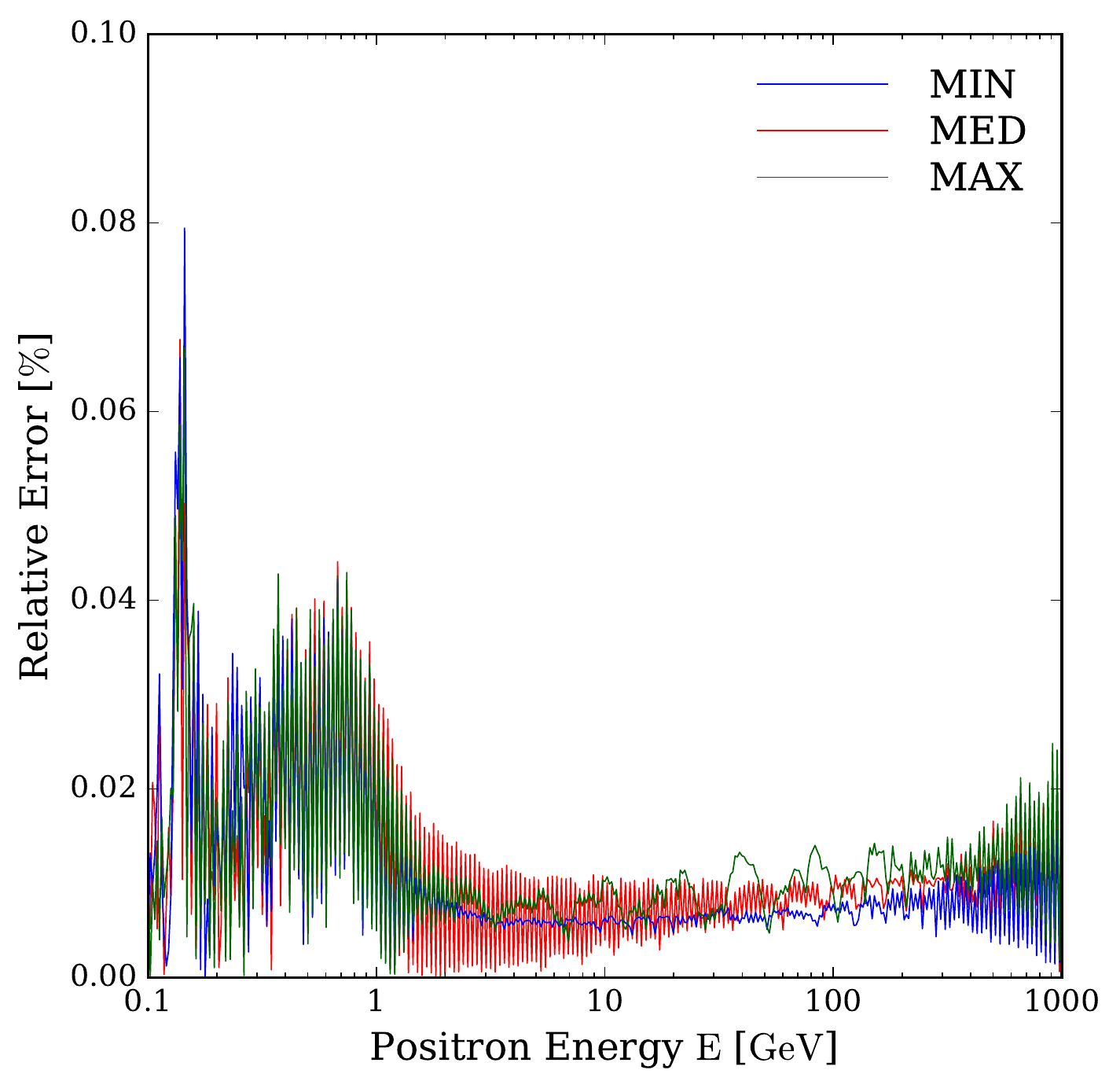}
\caption{{\bfseries Left panel:} IS secondary positron flux (multiplied by $E^{3.3}$) in the high-energy approximation scheme for the \MED\ model. {\bfseries Right panel:} relative error using the pinching method for secondary positrons.}
\label{fig:sec_flux}
\end{center}
\end{figure*}

\begin{figure*}[h!]
\begin{center}
\includegraphics[width=0.47\textwidth]{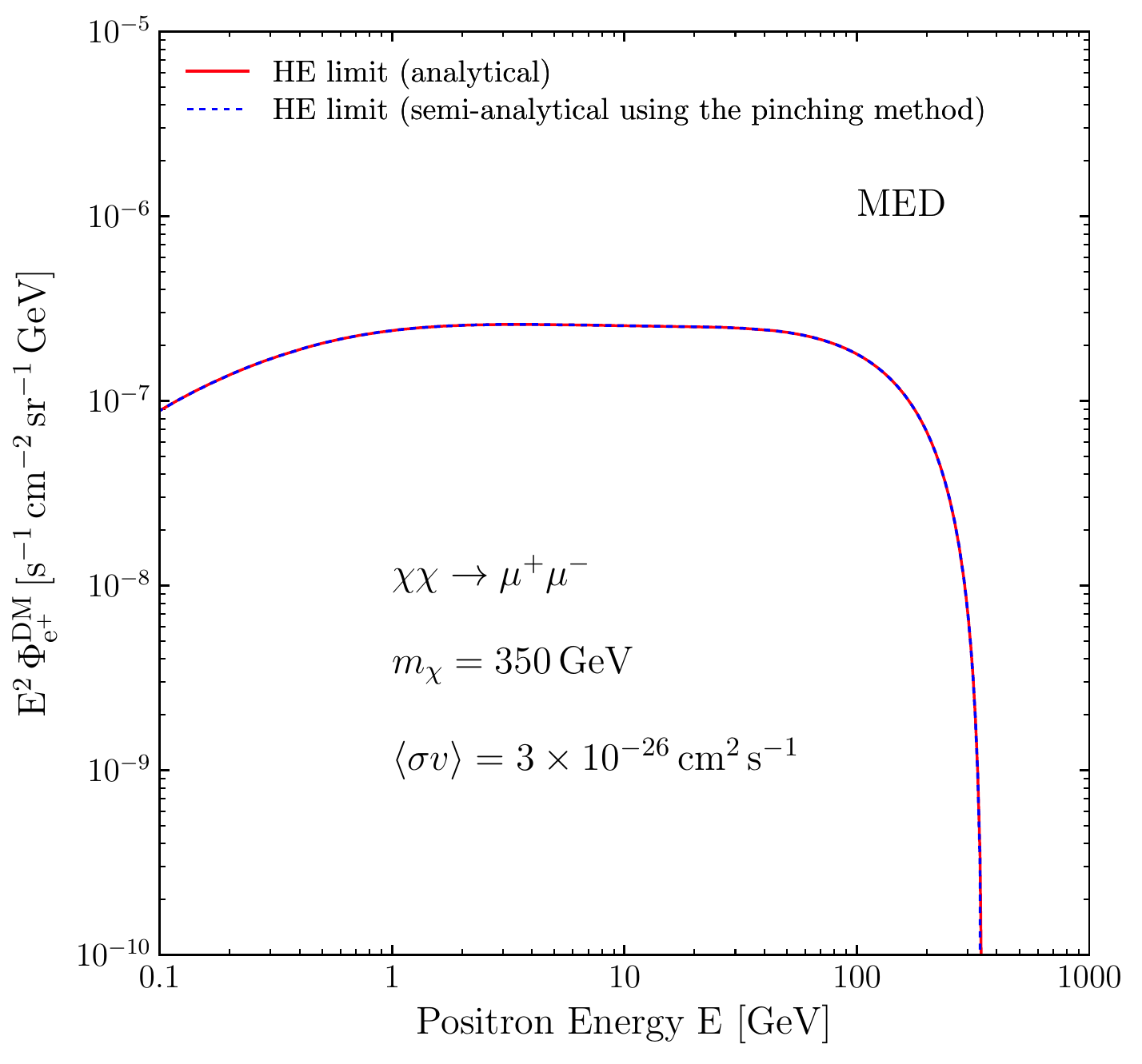}
\hspace{1cm}
\includegraphics[width=0.45\textwidth]{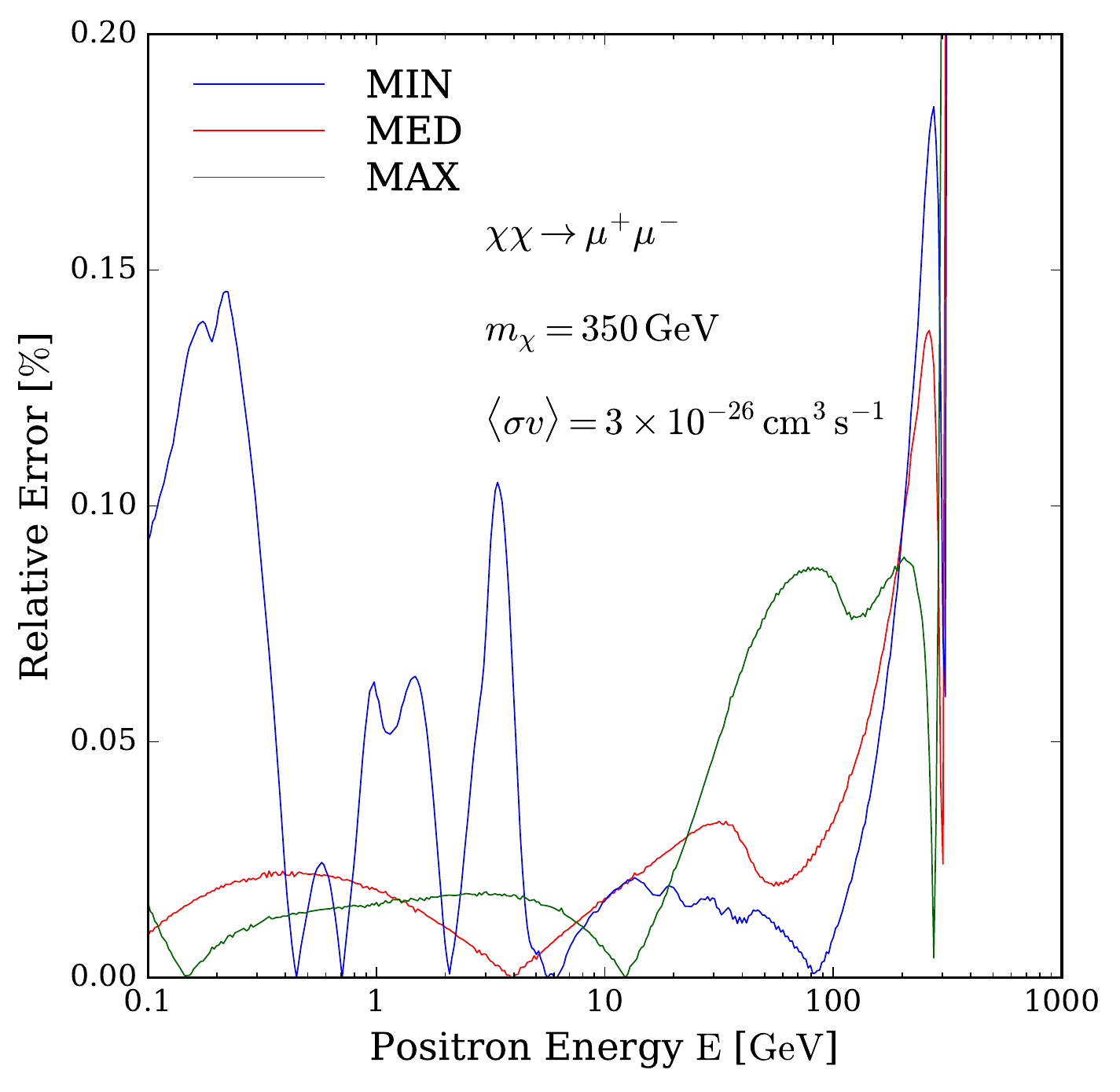}
\caption{{\bfseries Left panel:} IS positron flux (multiplied by $E^{2}$) produced by a {\rm 350 GeV} DM particle annihilating into $\mu^+ \mu^-$ pairs with $\langle \sigma v \rangle = 3 \times 10^{-26} \; {\rm cm}^3 \, {\rm s}^{-1}$ in the high-energy approximation scheme for the \MED\ model. {\bfseries Right panel:} relative error using the pinching method.}
\label{fig:prim_flux_mu}
\end{center}
\end{figure*}

\begin{figure*}[h!]
\begin{center}
\includegraphics[width=0.47\textwidth]{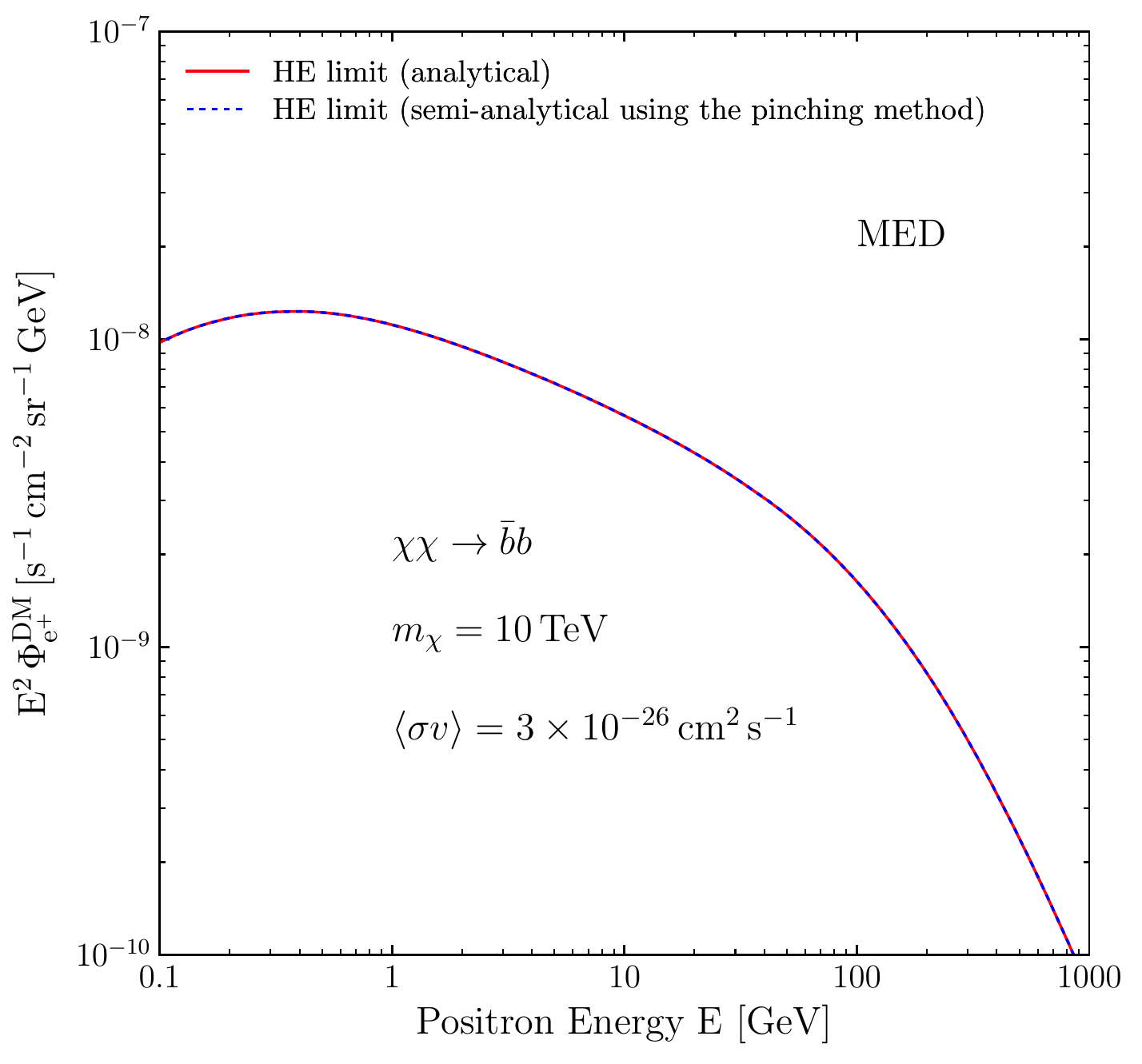}
\hspace{1cm}
\includegraphics[width=0.45\textwidth]{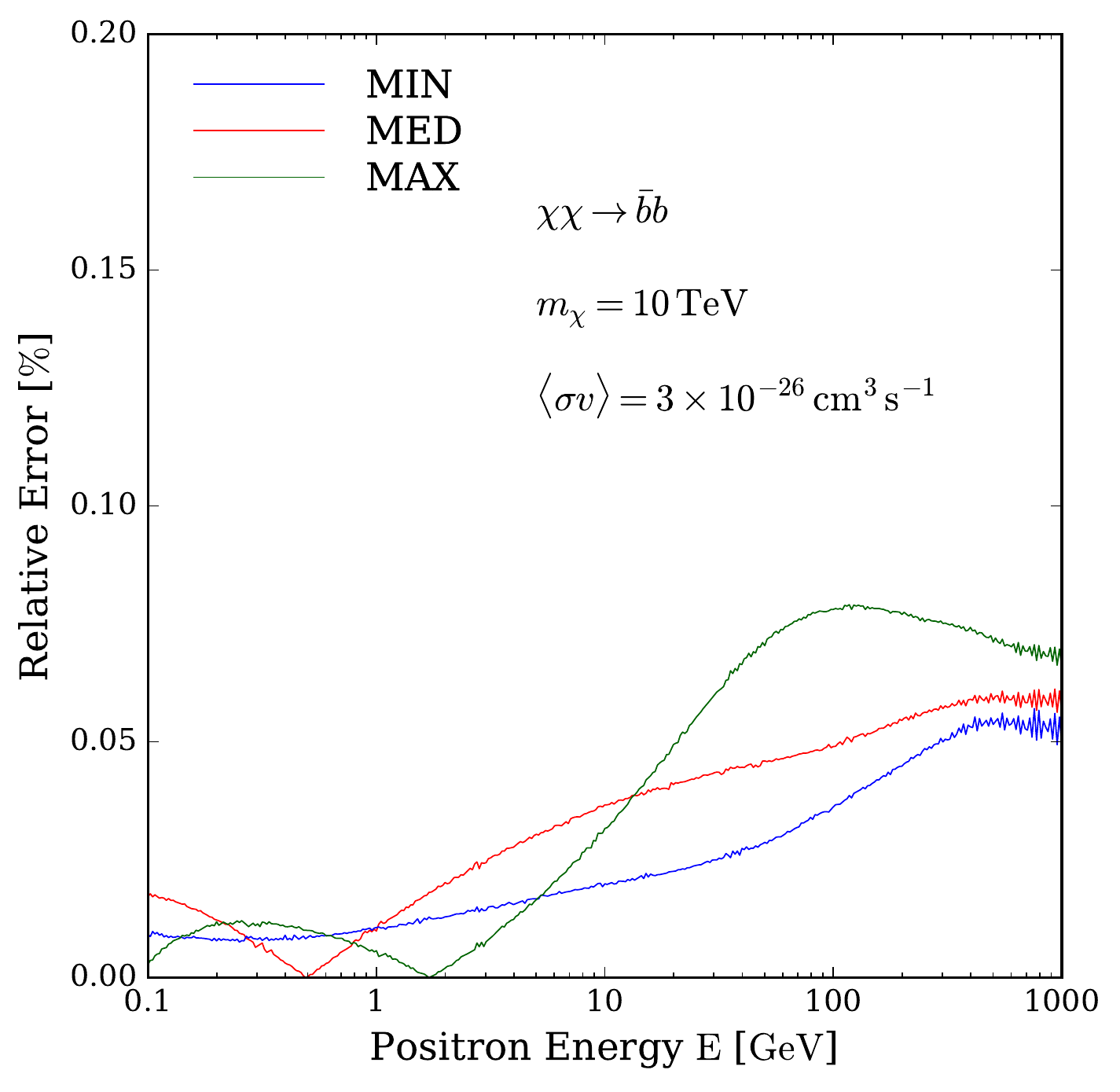}
\caption{Same as Fig.~\ref{fig:prim_flux_mu} with a {\rm 1 TeV} DM particle annihilating into $b\bar{b}$ pairs.}
\label{fig:prim_flux_b}
\end{center}
\end{figure*}

We represent in the left panel of of Fig.~\ref{fig:sec_flux} the secondary positron flux at the Earth computed in the high-energy approximation scheme with the \MED\ model. The red solid line represents the analytical solution whereas the blue dotted line represents the semi-analytical solution obtained when IC scattering and synchrotron energy losses are pinched in the Galactic disc. The relative error arising from the pinching method, is shown in the right panel of Figs.~\ref{fig:sec_flux} for \MIN\ (blue), \MED\ (red), and \MAX\ (green).
Furthermore, we plot in the left panels of Fig.~\ref{fig:prim_flux_mu} and Fig.~\ref{fig:prim_flux_b} both solutions for the primary positron flux produced respectively by a 350~GeV DM particle annihilating into $\mu^+ \mu^-$ and a 1~TeV DM particle annihilating into $b\bar{b}$. The cross section is taken to be $\langle \sigma v \rangle = 3 \times 10^{-26} \; {\rm cm}^3 \, {\rm s}^{-1}$. The relative error corresponding is represented in the right panels of Fig.~\ref{fig:prim_flux_mu} and Fig.~\ref{fig:prim_flux_b}.

For secondary positrons, this error is always kept below 0.1 \%. Our method is therefore very accurate at computing positrons produce by $p$ and He spallation onto the ISM. Regarding the primary contribution from DM annihilations, as long as the positron energy is well under the DM particle mass $m_\chi$, the error is also very small, always below 0.2\%. Close to $m_\chi$, the steep decrease of the positron flux (which eventually vanishes at $E\geq m_\chi$) induces a fast increase of the relative error. However, the error  is above 0.2\% only for energies at which the positron flux is highly suppressed. Therefore, we can safely consider that our technic will not introduce any sizeable bias in the analysis.

\vskip 0.1cm
Given its generality, the expression~(\ref{eq:xi_bar_r}) enables us to pinch IC and synchrotron energy losses in the Galactic disc regardless of their origins, i.e. whether they are secondary or primary CRs.
Thus, we can predict for the first time the electron flux at the Earth, including all propagation effects, using the semi-analytical resolution of the transport equation. In the following sections, we will apply our method to both secondary and primary CRs from DM annihilation to illustrate important differences with previous treatment. Our goal is now to recompute in the most accurate way propagation constraints from positron flux at the Earth and then reinvestigate the DM explanation of the excess.

%
%
%
%
%
%



\section{Implications for secondary positrons and the dark matter signal}
\label{sec:implication}
In this section, we compute the source term of secondary positrons with the up-to-date primary proton and helium fluxes. The interstellar flux of secondary positrons is derived semi-analytically with the pinching method presented in Sect.~\ref{sec:pinching}.
We then focus on the DM signal coming, as an illustration, from a 10 TeV WIMP annihilating into $\bar{b}b$ quark pairs.
These secondary and primary fluxes, computed including all the propagation processes, are compared with the ones derived from the high-energy approximation.

\subsection{Astrophysical background of secondary positrons}\label{sec:astrobkg}

Secondary positrons originate from the decay of pions, kaons, and delta baryons produced by inelastic collisions of primary CR protons and helium nuclei on the ISM. The injection rate of secondary positrons reads:
\beq
\label{eq:SecondarySourceTerm}
Q^{\rm II}(E, r, z) = 4 \pi \! \! \sum_{j = p, \rm He}  \sum_{i = \rm {H,He}}   n_i \int \! \! dE_j  \; \frac{d\sigma_{ji}}{dE}(E_j \rightarrow E) \; \Phi_j(E_j, r, z),
\eeq
where $n_i$ labels the atomic density of the nucleus $i$ in the ISM, $d\sigma_{ji}/dE$ indicates the positron differential production cross section, and $\Phi_j$ stands  for the CR proton and helium fluxes. We use the parameterisation of the proton-proton interaction differential cross section derived by\citetads{2006ApJ...647..692K}. For proton-helium interactions, we take the parameterisation from \citetads{2007NIMPB.254..187N}. To obtain the proton and helium fluxes everywhere in the Galaxy, we apply the retro-propagation method introduced by\citetads{2001ApJ...555..585M}, which requires as an input the TOA flux. This work is based on the latest measurements by \AMS\citepads{2015PhRvL.114q1103A,2015PhRvL.115u1101A} and \CREAM\citepads{2011ApJ...728..122Y}. The proton and helium fluxes are fitted using a model introduced in\citetads{2015PhRvL.114q1103A,2015PhRvL.115u1101A}, where a single power law in rigidity $R^{\gamma}$ exhibits a smooth transition  to $R^{\gamma+\Delta \gamma}$ above the rigidity $R_\mathrm{b}$. The smoothness of the spectral index transition is described by the parameter $s$. An additional effective parameter  $\alpha$ is used to fit the low-rigidity part of the proton flux. The interstellar (IS) primary fluxes can be described as follows:
\begin{equation}
\Phi^{\rm IS}_{p}(R) = C \, \beta \, \left(1-e^{\alpha R}\right) R^{\gamma} \left[1+\left(\frac{R}{R_\mathrm{b}}\right)^{\Delta \gamma / s}\right]^s,
\label{eq:protonFluxModel}
\end{equation}
and
\begin{equation}
\Phi^{\rm IS}_{\rm He}(R) = C \, \beta \, R^{\gamma} \left[1+\left(\frac{R}{R_\mathrm{b}}\right)^{\Delta \gamma / s }\right]^s,{}
\label{eq:heliumFluxModel}
\end{equation}
with $\beta$ the particle velocity. The force-field approximation\citepads{1971JGR....76..221F} is used to obtain the relation between $\Phi_{\rm IS}$ and $\Phi_{\rm TOA}$, \ie\ respectively the IS and TOA fluxes. The value $\phi_\mathrm{F} = 724 \; \rm MV$ determined by\citetads{2016A&A...591A..94G} is used hereafter unless explicitly stated.
%
\begin{figure*}[ht!]
\begin{center}
\includegraphics[width=0.39\textwidth]{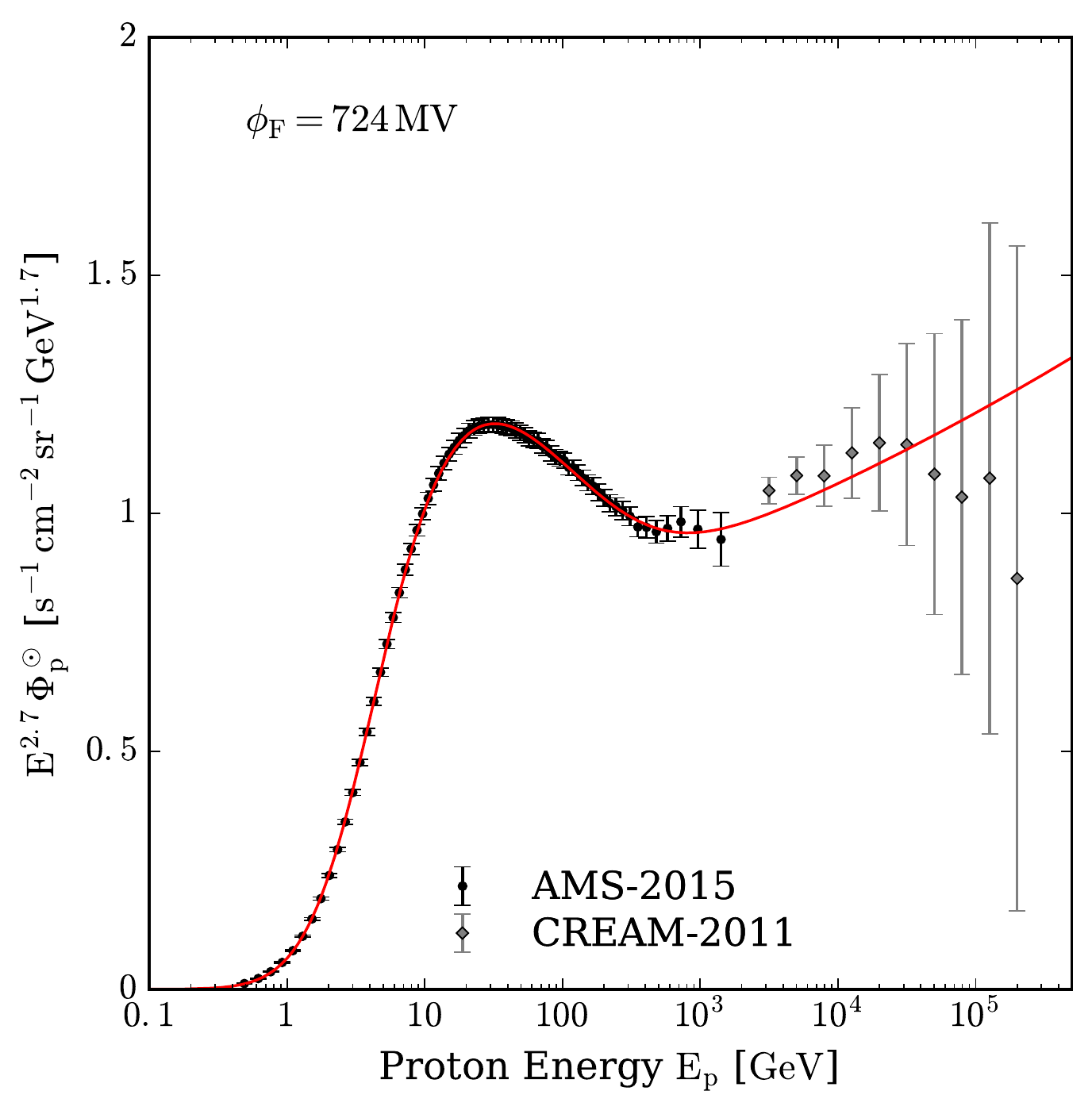}
\hspace{1cm}
\includegraphics[width=0.4\textwidth]{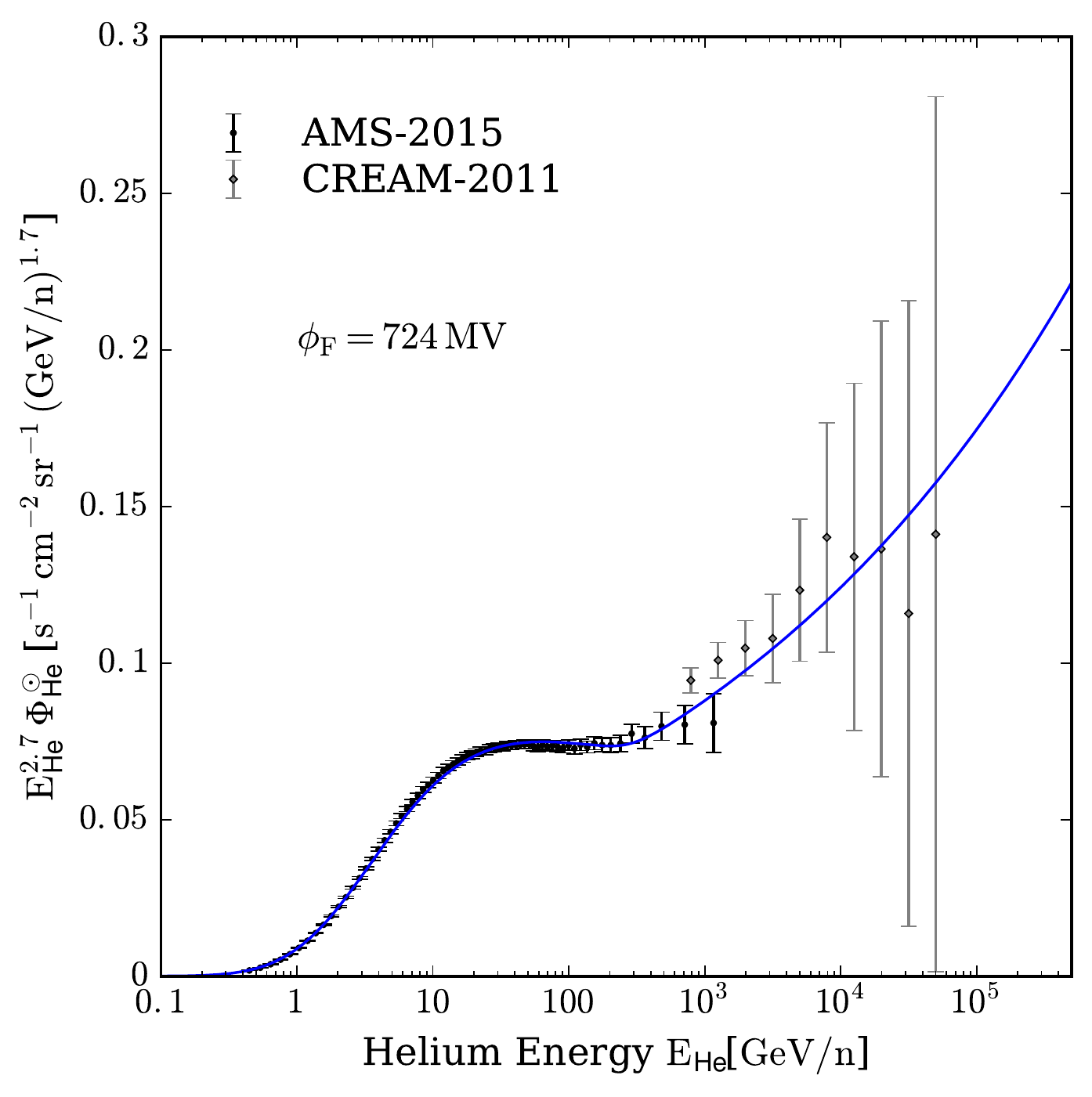}
\caption{
Proton (\textbf{left panel}) and helium (\textbf{right panel}) fluxes (multiplied by $E^{2.7}$) as a function of kinetic energy for \AMS\citepads{2015PhRvL.114q1103A,2015PhRvL.115u1101A} and \CREAM\citepads{2011ApJ...728..122Y} data. The red and blue curves are the fitted proton and helium fluxes corresponding respectively to Eqs.~(\ref{eq:protonFluxModel}) and (\ref{eq:heliumFluxModel}). We use the value from\citetads{2016A&A...591A..94G} of $724$ MV for the Fisk potential $\phi_\mathrm{F}$.}
\label{fig:BESSAMSCREAM}
\end{center}
\end{figure*}
%
This model has been fitted to the measured fluxes, as it is shown in Fig.~\ref{fig:BESSAMSCREAM}, yielding the parameter values reported in Table~\ref{tab:parameterProton}.
%
\begin{table*}[h!]
\begin{center}
\caption{
{\small  Values of the proton and helium flux parameters resulting from a fit to the \AMS\citepads{2015PhRvL.114q1103A,2015PhRvL.115u1101A} and \CREAM\citepads{2011ApJ...728..122Y} data assuming $\phi_\mathrm{F} = 724 \; \rm MV$.}
}\begin{tabular}{ccccccc}
\hline
\hline
\CellTop
{} & {$C$ [m$^{-2}$ s$^{-1}$ sr$^{-1}$ GV$^{-1}$]} & {$\alpha$ [GV$^{-1}$]} & $\gamma$ & $R_\mathrm{b}$ [GV] & $\Delta \gamma$ & $s$ \\
\hline
\CellTop
{Proton} & $(2.71 \pm 0.02) \times 10^{4}$ & $-0.512 \pm 0.012$ & $-2.88 \pm 0.01$ & $424 \pm 158$ & $0.242 \pm 0.056$ & $0.156 \pm 0.072$ \\
\hline
{Helium} & $(3.56 \pm 0.04) \times 10^{3}$ & - & $-2.77 \pm 0.01$ & $543 \pm 163$ & $0.213 \pm 0.045$ & $0.047 \pm 0.018$ \\
\hline
\end{tabular}
\label{tab:parameterProton}
\end{center}
\end{table*}
%

\begin{figure*}[h!]
\begin{center}
\includegraphics[width=0.42\textwidth]{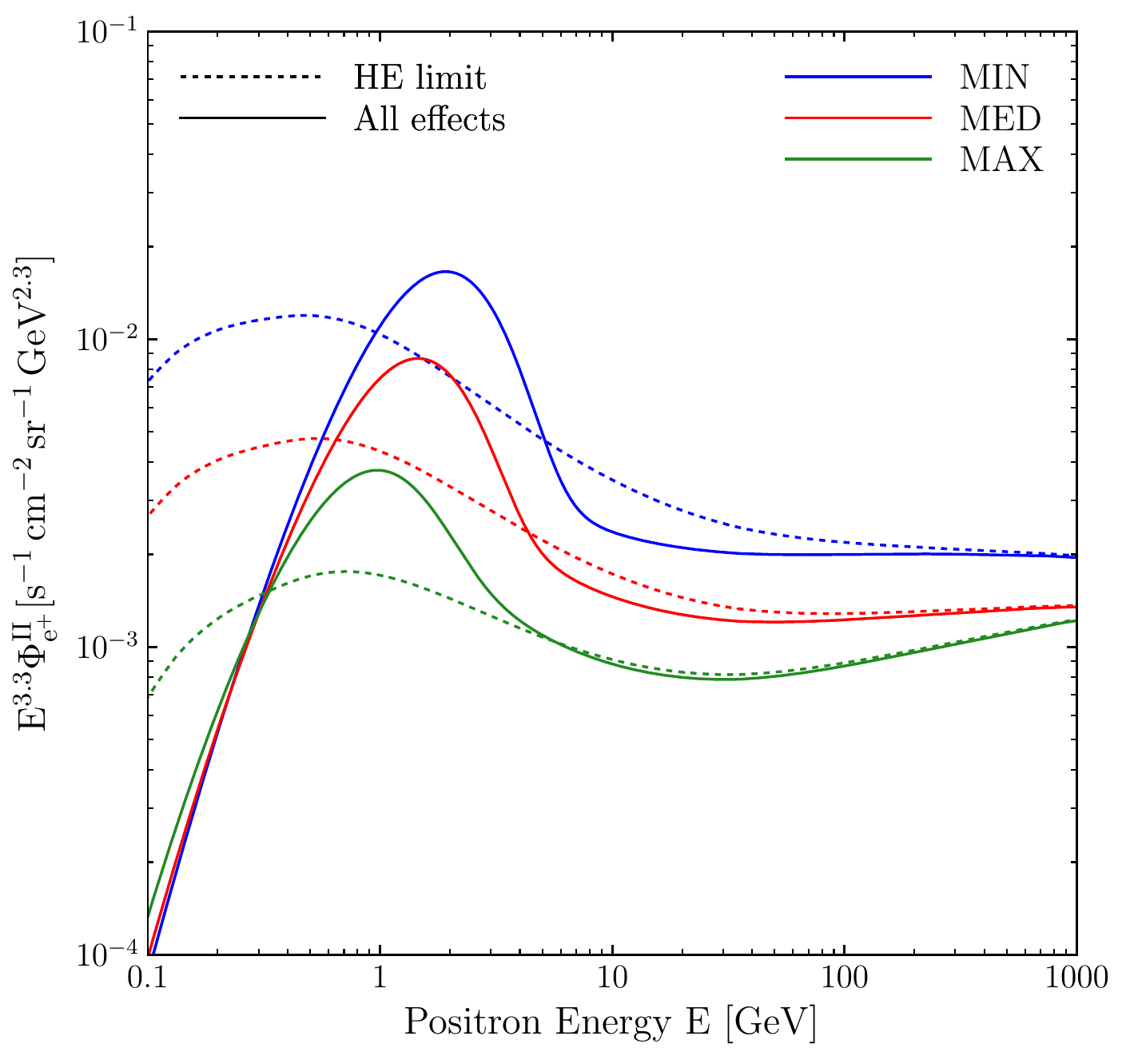}
\hspace{1cm}
\includegraphics[width=0.4\textwidth]{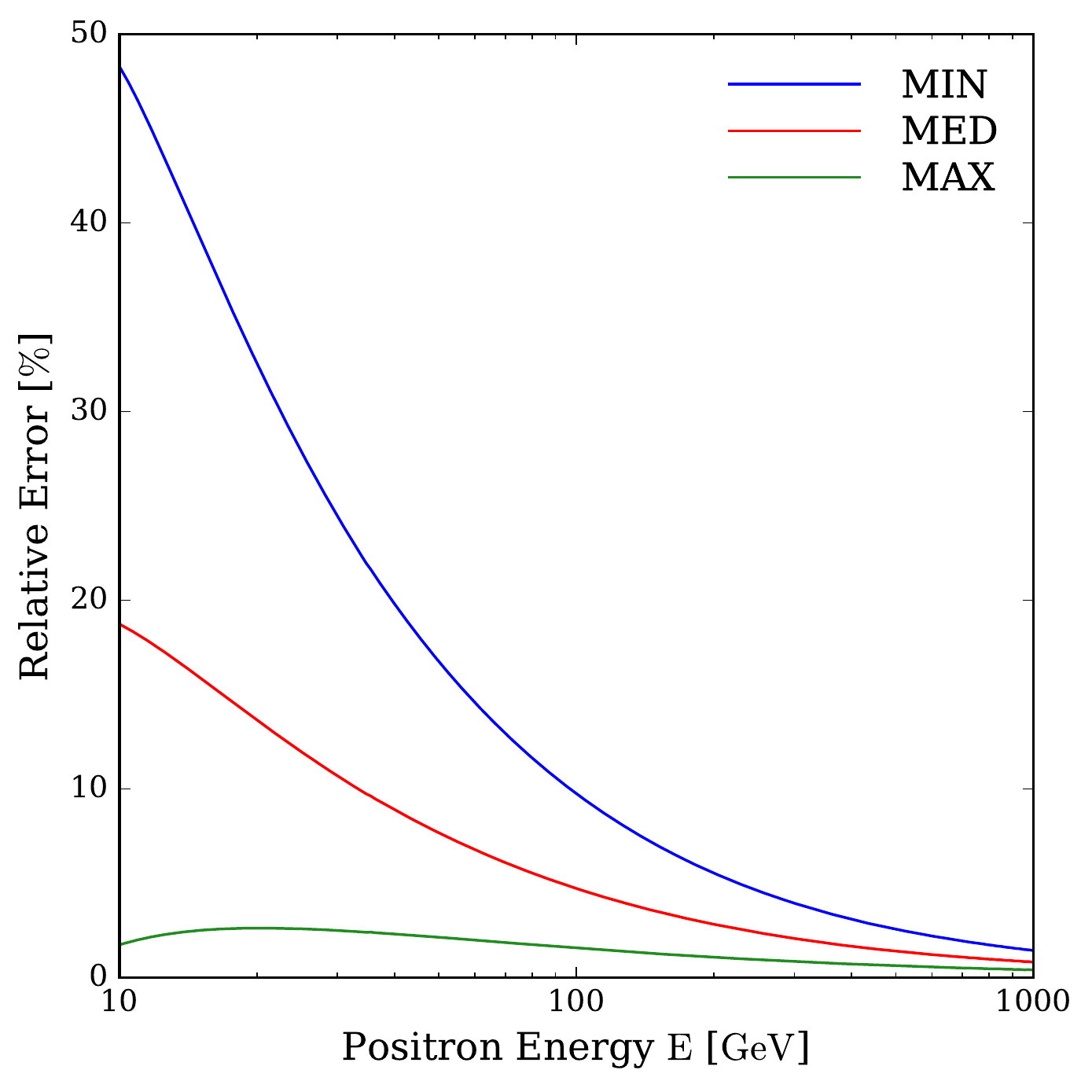}
\caption{
\textbf{Left panel:} interstellar flux (multiplied by $E^{3.3}$) of secondary positrons computed with all propagation effects ($\Phi^{\rm II}$, solid lines) and with the high-energy approximation ($\Phi^{\rm II}_{\rm HE}$, dotted lines) for the \MIN\ (blue), \MED\ (red), and \MAX\ (green) models. \textbf{Right panel:} relative error $(\Phi^{\rm II}_{\rm HE} - \Phi^{\rm II}) / \Phi^{\rm II}$ above 10~GeV of the high-energy approximation for secondary positrons compared to the exact result.}
\label{fig:sec_flux_MIN_MED_MAX}
\end{center}
\end{figure*}

\vskip 0.1cm
The interstellar flux of secondary positrons, computed with the pinching method including all propagation effects, is represented in the left panel of Fig.~\ref{fig:sec_flux_MIN_MED_MAX} by the solid lines for \MIN\ (blue), \MED\ (red), and \MAX\ (green).
The high-energy approximation, where only diffusion and halo energy losses are taken into account, is featured by the dotted lines.
It is henceforth possible to assess the error made when applying the high-energy approximation often used in the literature to compute the positron flux above 10~GeV. This error is defined as $( \Phi^{\rm II}_{\rm HE} - \Phi^{\rm II}) / \Phi^{\rm II}$ where the index $\rm HE$ stands for high energy. This quantity is plotted in the right panel of Fig.~\ref{fig:sec_flux_MIN_MED_MAX}, and a few numerical values are displayed in Table~\ref{tab:error_HE_approx_II}.
As already noticed by\citetads{2009A&A...501..821D}, the high-energy approximation tends to largely underestimate the amount of positrons below $5$~GeV. Interestingly, we find on the other hand that above that value, the high-energy approximation overshoots the exact result.
Indeed, although convection and disc energy losses are subdominant with respect to halo energy losses and space diffusion, they still have a sizeable effect and tend to reduce the positron flux above 10~GeV.
Moreover, the relative error strongly depends on the propagation parameters, the maximum value beeing reached for the \MIN\ configuration. This can be understood by the fact that the convection velocity decreases along the sequence \MIN, \MED, \MAX. Therefore, we observe that the discrepancy with the high-energy approximation increases with higher values of the convection.
%
%
%
\begin{table*}[h!]
\begin{center}
\caption{
Typical values of the relative error $( \Phi^{\rm II}_{\rm HE} - \Phi^{\rm II}) / \Phi^{\rm II}$ (\%) of the high-energy approximation for secondary positrons compared to the exact result.}
\begin{tabular}{cccccc}
\hline
\hline
\CellTop
Positron energy (GeV) 			& 10				& 50		& 100		& 500 		& 1000 \\
\hline
\CellTop
\MIN  						& 48					& 17			&  9.7 			& 2.5 			& 1.4 \\
\MED  						& 19					& 7.7			&  4.7	 		& 1.4	 		& 0.8 \\
\MAX  						& 1.7					& 2.0			&  1.5			& 0.6	 		& 0.4\\
\hline
\end{tabular}
\label{tab:error_HE_approx_II}
\end{center}
\end{table*}
%

\subsection{Primary positrons from the annihilation of dark matter particles}\label{prim:DM_species}

The source term of positrons produced by the annihilation of DM particles reads
\beq
\label{eq:source_DM}
Q^{\rm DM}(E, \vec{x}) =\eta \, \langle \sigma v \rangle \, \frac{\rho^2_{\chi}(\vec{x})}{m_{\chi}^2} \,   \left\{ g(E) \equiv \sum_i^N b_i \left. \frac{dN}{dE} \right|_i \right\},
\eeq
where $m_{\chi}$ is the DM particle mass and $\langle \sigma v \rangle$ its average annihilating cross section. The value of $\eta$ depends on whether the DM particle is Majorana-type ($\eta = 1/2$) or Dirac-type ($\eta = 1/4$).
We use the DM density profile introduced by\citetads{1997ApJ...490..493N}, hereafter denoted NFW, with the local DM density $\rho_\odot = 0.3 \; \rm GeV \, cm^{-3}$\citepads{2012ApJ...756...89B}.
The energy distribution of positrons $g(E)$ at the source is obtained by summing over the individual contributions $dN/dE |_i$ for each annihilating channel $i$ weighted by the branching ratio $b_i$.
The individual energy distributions $dN/dE |_i$ are computed with the micrOMEGAs\_3.6 package\citepads{2011CoPhC.182..842B,2014CoPhC.185..960B}.

\begin{figure*}[h!]
\begin{center}
\includegraphics[width=0.415\textwidth]{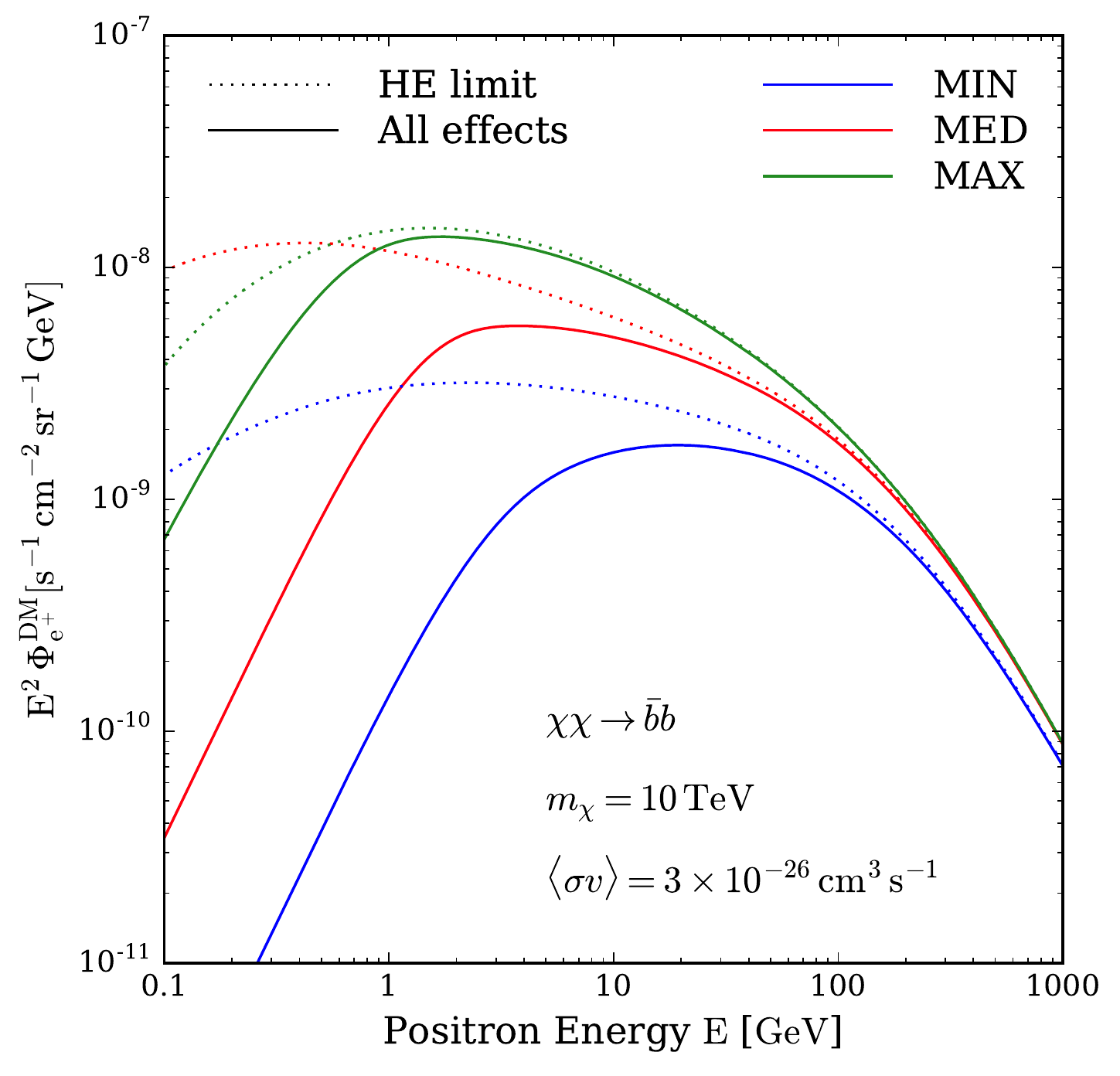}
\hspace{1cm}
\includegraphics[width=0.4\textwidth]{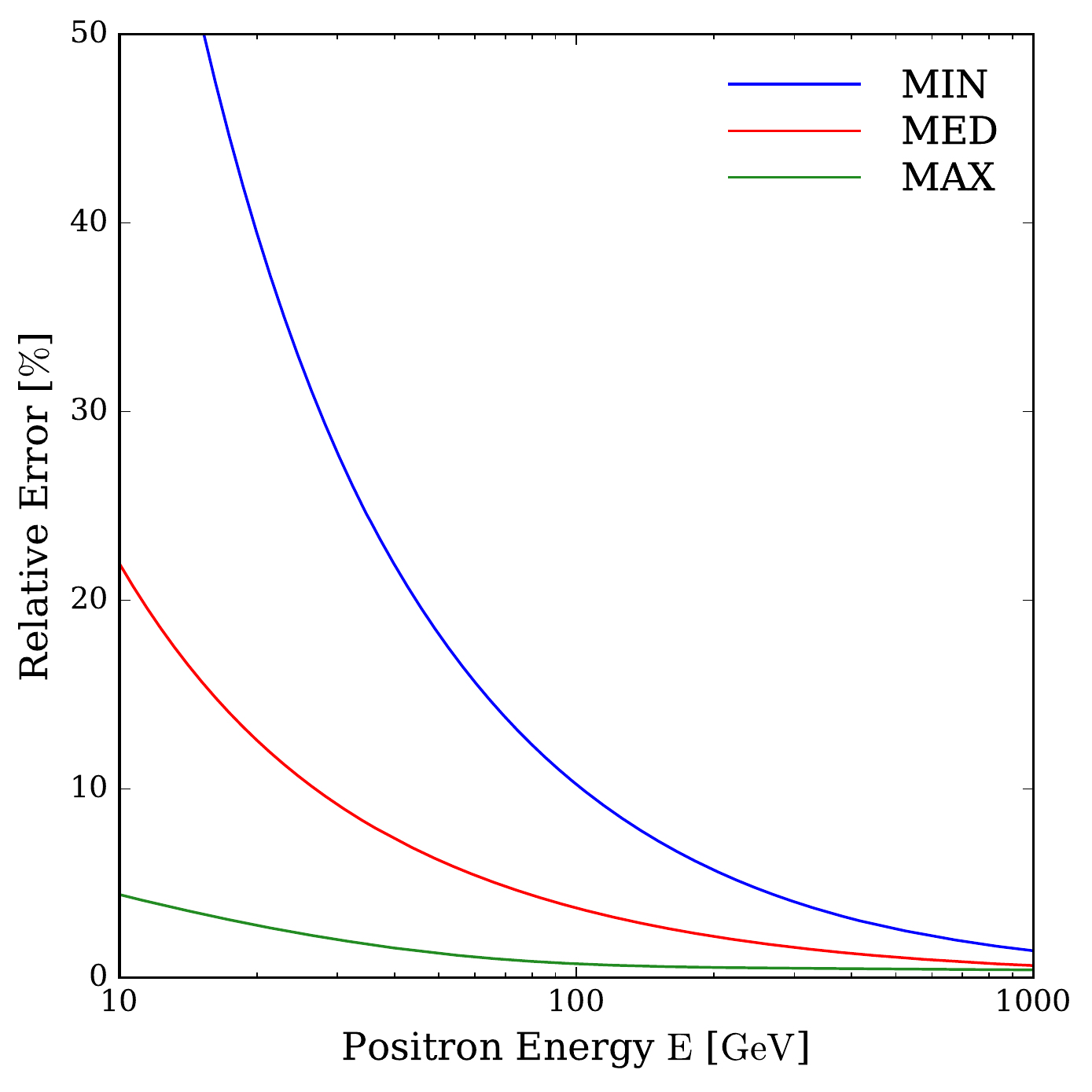}
\caption{
{\bfseries Left panel:}
interstellar flux (multiplied by $E^2$) of primary positrons computed with all propagation effects ($\Phi^{\rm DM}$, solid lines) and with the high-energy approximation ($\Phi^{\rm DM}_{\rm HE}$, dotted lines) for a 10~TeV WIMP annihilating into $\bar{b}b$ pairs with $\langle \sigma v \rangle = $ $3 \times 10^{-26}$ {\rm cm}$^3$ {\rm s}$^{-1}$, for the \MIN\ (blue), \MED\ (red), and \MAX\ (green) models.
{\bfseries Right panel:} relative error $( \Phi^{\rm DM}_{\rm HE} - \Phi^{\rm DM}) / \Phi^{\rm DM}$ above 10~GeV of the high-energy approximation for primary positrons compared to the exact result.}
\label{fig:DM_flux_MIN_MED_MAX}
\end{center}
\end{figure*}

\vskip 0.1cm
For illustrative purposes, we consider throughout this section a Majorana-type DM species with a mass $m_\chi$ of $10$ TeV annihilating into $\bar{b}b$ quark pairs with the thermal cross section $\langle \sigma v \rangle = 3 \times 10^{-26} \; \rm cm^3 \, s^{-1}$.
The IS flux of primary positrons computed with all propagation processes taken into account is featured by the solid lines in the left panel of Fig.~\ref{fig:DM_flux_MIN_MED_MAX}.
The high-energy approximation corresponds to the dotted lines.
The relative error $( \Phi^{\rm DM}_{\rm HE} - \Phi^{\rm DM}) / \Phi^{\rm DM}$ is plotted in the right panel whereas a few numerical values are displayed in Table~\ref{tab:error_HE_approx_I}. We notice differences in the magnitude of this error, depending on the CR propagation configuration. We attribute them to the different values of the convective velocity \Vc. Actually, positrons produced by DM annihilating throughout the MH are more sensitive to convection than secondary positrons, which originate from the Galactic disc. As a consequence, the error associated to the high-energy approximation tends to be larger for primary positrons than for secondary ones. In the former case, it is significantly large in the \MIN\ model for which \Vc is the highest.
%
%
%
\begin{table*}[h!]
\begin{center}
\caption{
Typical values of the relative error $( \Phi^{\rm DM}_{\rm HE} - \Phi^{\rm DM}) / \Phi^{\rm DM}$ of the high-energy approximation for primary positrons compared to the exact result.}
\begin{tabular}{cccccc}
\hline
\hline
\CellTop
Positron energy (GeV)  			& 10				& 50		& 100		& 500		& 1000 \\
\hline
\CellTop
\MIN  						& 74					& 18			&  10 			&  2.6			&  1.4 \\
\MED  						& 22					& 6.2			& 3.8 	 		& 1.1	 		& 0.6 \\
\MAX  						& 4.4					& 1.3			& 0.7 			& 0.5	 		& 0.4\\
\hline
\end{tabular}
\label{tab:error_HE_approx_I}
\end{center}
\end{table*}
%

\vskip 0.1cm
In summary, we have computed the flux of positrons including all the propagation effects for the secondary component as well as the DM signal.
We have shown that low-energy effects modify drastically the shape of the positron spectrum. In addition, these effects could have a sizeable importance above 10 GeV, in contrast to what has been assumed in the literature. At 10~GeV, they modify the prediction up to 48\% for the secondary component, and up to 74\% for the DM signal, in the \MIN\ configuration.
Therefore, neglecting the low-energy CR propagation processes could lead to misleading interpretations when attempting to compare the theoretical predictions to the high-accuracy data provided by the \AMS\ collaboration.
All the results presented in the following of this paper are obtained using the pinching method to solve the \textit{full} transport equation Eq.~(\ref{eq:full_transport}).



\section{Constraining propagation parameters with A{\sc ms}-02 data\label{sec:prop_param}}

\subsection{Secondary positrons and propagation models}

Secondary cosmic rays are often regarded as a powerful observable to constrain the propagation scenario. Instead of relying on unknown source modelisation, their source term is determined by primary particles, for which precise measurements are available, therefore allowing to disentangle more easily propagation from injection effects~\footnote{Potential contamination by non-negligible primary component could spoil such ability, see e.g. \citetads{2015A&A...580A...9G} for a discussion.}. This is the case of the boron flux, commonly divided by the carbon flux, so that the B/C ratio no longer depends on the carbon injection assumptions. Secondary isotopes of helium and hydrogen (see for example\citeads{2012A&A...539A..88C}) and subFe/Fe ratio are also used for that purpose, and lead to similar understanding of CR propagation in our Galaxy.

Until the discovery of a high-energy excess, positrons have been thought for a long time as being pure secondary particles. Although its secondary component tends to be forgotten behind the excitement of such discovery, they still carry a wealth of information on propagation properties. In fact, as noticed in\citeads{2014PhRvD..90h1301L}, in many propagation models compatible with the B/C ratio (especially those with a small halo size), pure secondary predictions of the flux at the lowest energies (typically below 4 GeV) are not in deficit but rather {\em in excess} with respect to measurements. This observation has been shown to yield a useful complementary constraint on the propagation parameters. Indeed, since the flux of secondary positrons scales as the ratio of the production volume over the diffusion one, leading to $1/\sqrt{K_{0}}$ dependency, the well-known degeneracy $K_0/L$ introduced by secondary-to-primary ratio studies can be lifted. However, in order to use this complementarity, one needs in practice to be able to compute accurately the positron spectrum at the lowest energies, despite the presence of DR, convection, and disc energy losses. In\citeads{2014PhRvD..90h1301L}, a qualitative trick was used: it was argued that the inclusion of DR would lead to the formation of a bump around 1 GeV which tends to {\em increase} the flux with respect to cases in which it is neglected, thus leading to a predicted flux in excess of the data.
We have shown in Sect.~\ref{sec:implication} that the competition between DR, convection, and disc energy losses, tends to the formation of such a bump around 2 GeV. However, already above 5 GeV, there might be parts of the $\{\Vc,\Va\}$ parameter space that actually lead to a {\em decrease} of the flux. This is particularly pronounced in the \MIN\ model as shown in Fig.~\ref{fig:sec_flux_MIN_MED_MAX}. We will therefore recompute the constraints of\citeads{2014PhRvD..90h1301L} with i) our full resolution method at low energy, and ii) updated fluxes measured by \AMS. This will lead to more robust and more stringent constraints on the propagation parameters.

\subsection{Skimming method for the propagation models}

We compute the secondary positron flux for the 1,623 propagation parameter sets selected by the B/C ratio analysis of\citetads{2001ApJ...555..585M}. These parameters are sorted from a uniform linear grid in the propagation parameter space, namely ($\delta$, $K_0$, $L$, $\Vc$, $\Va$), and are in agreement with the \HEA B/C ratio within 3 standard deviations. The secondary positrons are calculated including all the effects describe in Sect.~\ref{sec:pinching} and recalled hereafter: diffusion, convection, reacceleration, high-energy losses (IC, synchrotron), low-energy losses (adiabatic, ionisation, coulombic, bremsstrahlung), retro-propagation of the proton and helium fluxes, annihilation, and solar modulation. One may worry that our constraints highly depend on solar modulation modelisation. Although no extensive study of solar modulation for positrons during the period for which \AMS has been taking data is available, this modulation is commonly assumed to affect equally particles of same rigidity and same sign of charge. This assumption will soon be tested by the forthcoming \AMS measurements of the variations of the positron-to-proton ratio over the last solar cycle. Therefore, within the force-field approximation, we can rely on studies of the proton solar modulation such as in\citetads{2016A&A...591A..94G} and make use of the Fisk potential derived there. In a conservative approach, we lower the secondary prediction as much as possible using the $3 \sigma$ {\em highest} Fisk potential which was found by\citetads{2016A&A...591A..94G} to be 830~MV. The constraints derived with this high value might not be optimal. They already provide quite strong conclusions as discussed in the following sections.

\vskip 0.1cm
In order to quantify, for a given propagation model, the deviations of the predicted flux from the data, and any potential overshooting, we follow the criterion advocated in\citetads{2014PhRvD..90h1301L}, and first define, for each energy bin, the quantity
\beq
 Z_i=\frac{\Phi^{\rm II}_{e^+}(E_i)-\Phi^{\rm{data}}(E_i)}{\sigma^{\rm{data}}(E_i)}\;,
\eeq
where $\Phi^{\rm II}_{e^+}(E_i)$ is the predicted secondary positron flux in a given energy bin, $\Phi^{\rm{data}}(E_i)$ is the corresponding experimental flux, and $\sigma^{\rm{data}}(E_i)$ its experimental uncertainty. A propagation model is allowed provided that $Z_i$ does not exceed 3 whatever the energy bin. In other words, for selected models, we allow predictions to overshoot the data by at most 3 standard deviations in each energy bin. Note that, unlike\citeads{2014PhRvD..90h1301L}, we do not combine the values of $Z_i$ at different energies into a single statistical test. To do so, one would need to know correlations of experimental uncertainties between differents energies, but those are not provided by the \AMS collaboration. One could assume uncorrelated uncertainties, but this would be only true for the statistical ones. We therefore consider bins separetely, making our test a conservative choice over which there could be room for some improvement.

\subsection{Results and discussion}

\begin{figure*}[ht!]
\begin{center}
\includegraphics[width=0.99\columnwidth]{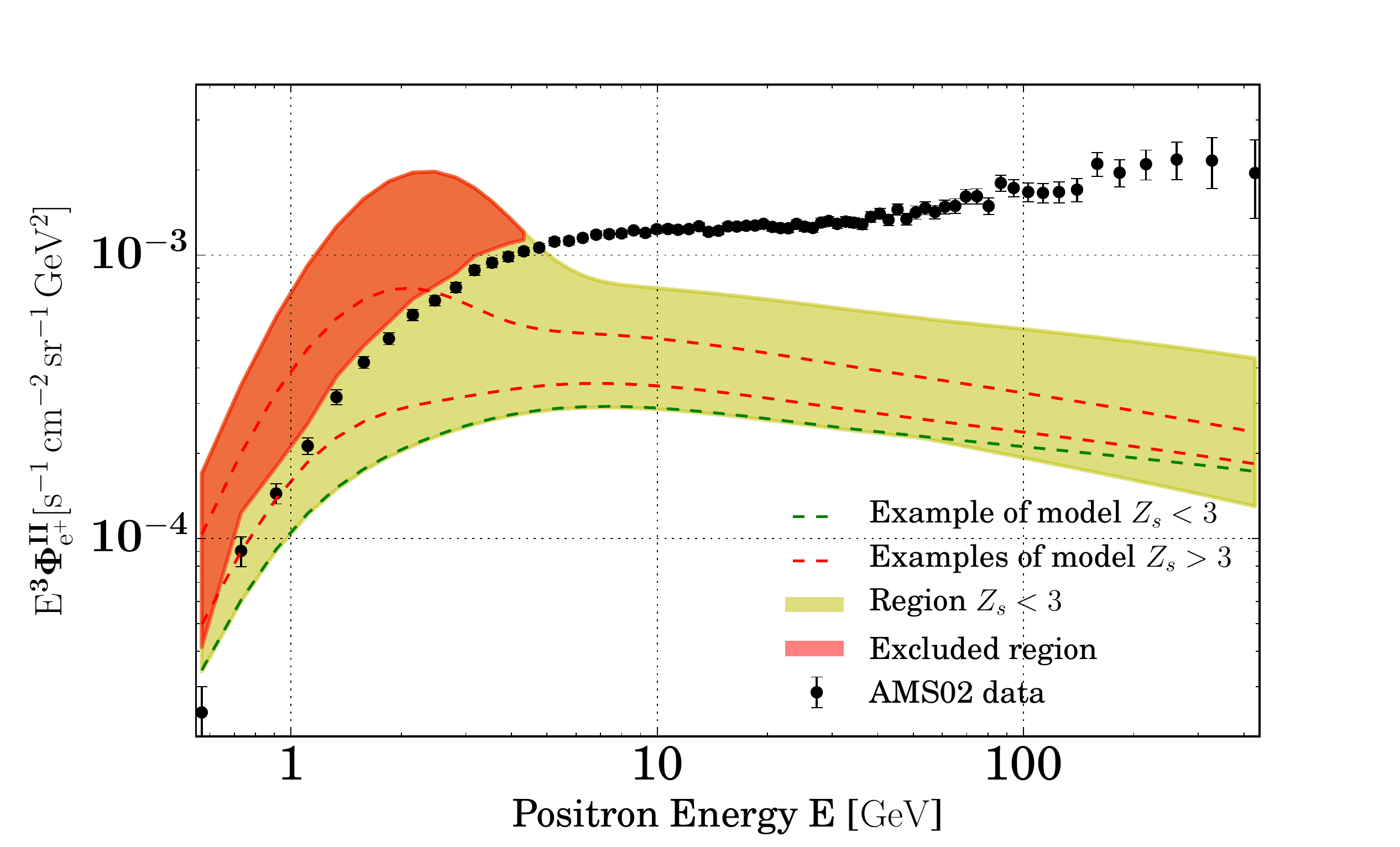}
\includegraphics[width=0.99\columnwidth]{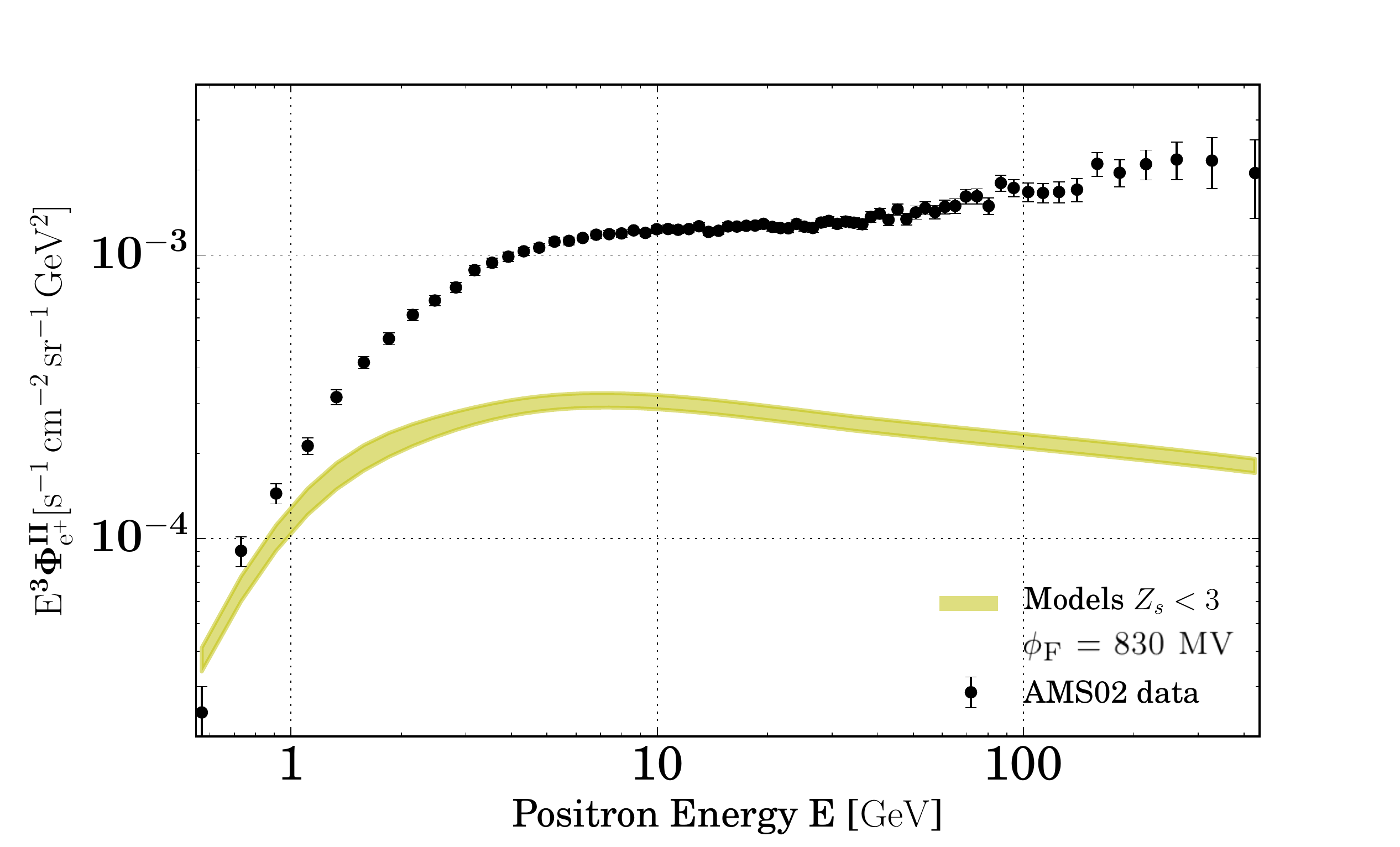}
\caption{\textbf{Left panel:} positron flux (multiplied by $E^3$) from the \AMS data compared to the envelope of the 1,623 flux predictions for the secondary positrons (colored band). The red colored region is the domain which is crossed by all the excluded models. The dashed red lines show two examples of models that do not fulfill the constraints $Z_i<3$ for all energy bins $i$ (i.e. the Z-score constraint defined as $Z_{s} = {\max} \,(Z_i) <3$). The dashed green line represents an example of a model that fulfills the constraints $Z_i<3$ for all energy bins $i$. \textbf{Right panel:} positron flux (multiplied by $E^3$) from the \AMS data compared to the envelope of the 54 models fulfilling $Z_i<3$ for all energy bins $i$.}
\label{fig:ecremage}
\end{center}
\end{figure*}

\begin{figure*}[ht!]
\begin{center}
\includegraphics[width=0.99\columnwidth]{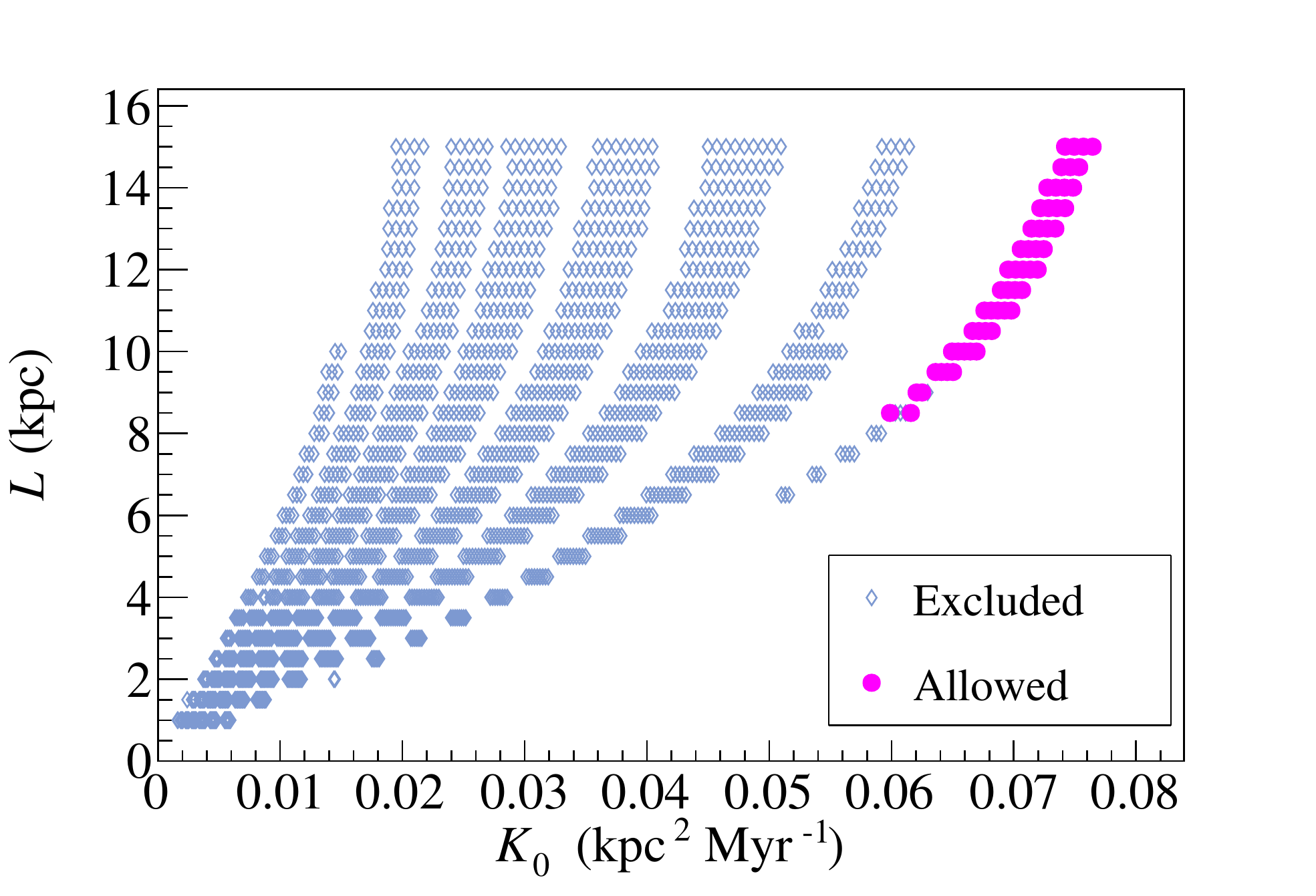}
\includegraphics[width=0.99\columnwidth]{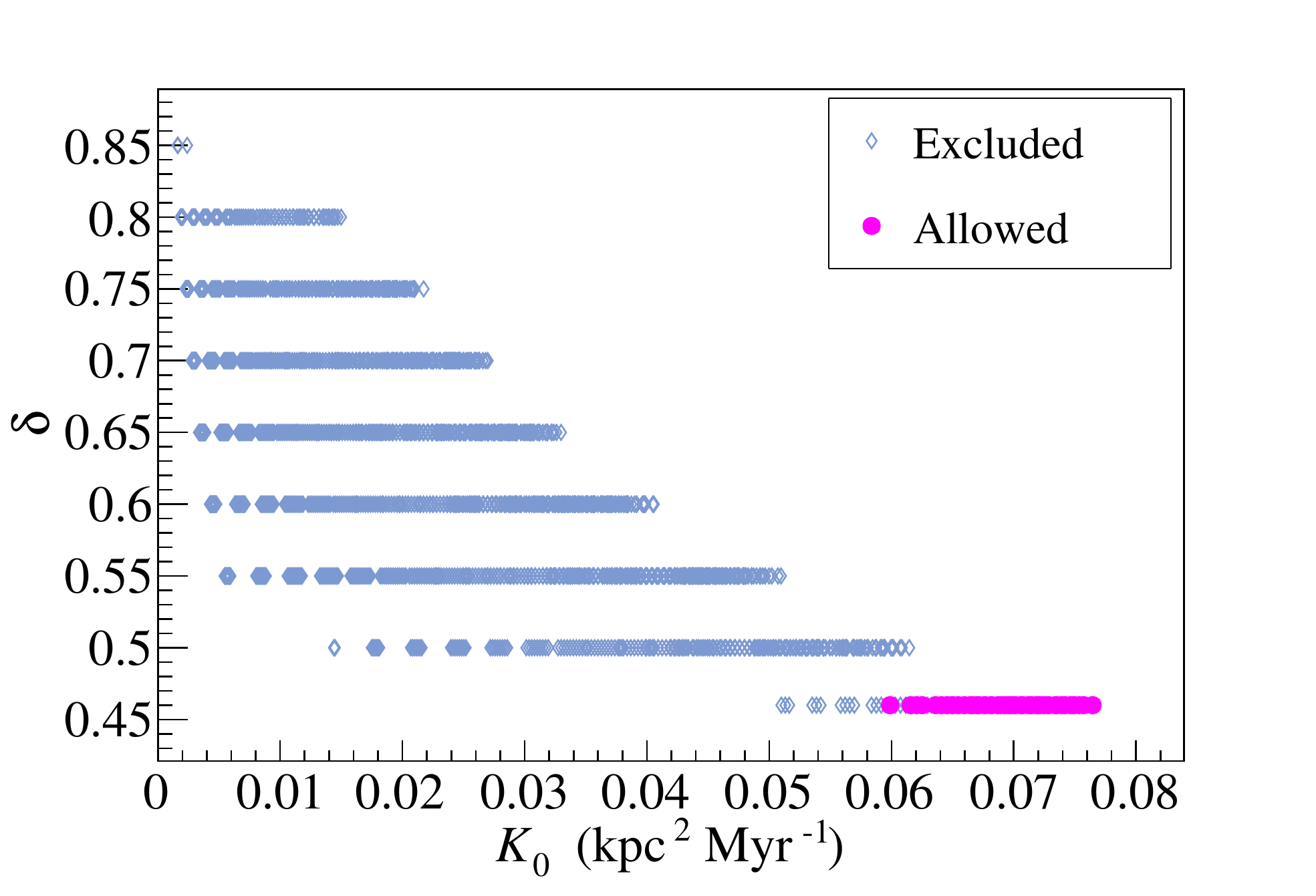}
\includegraphics[width=0.99\columnwidth]{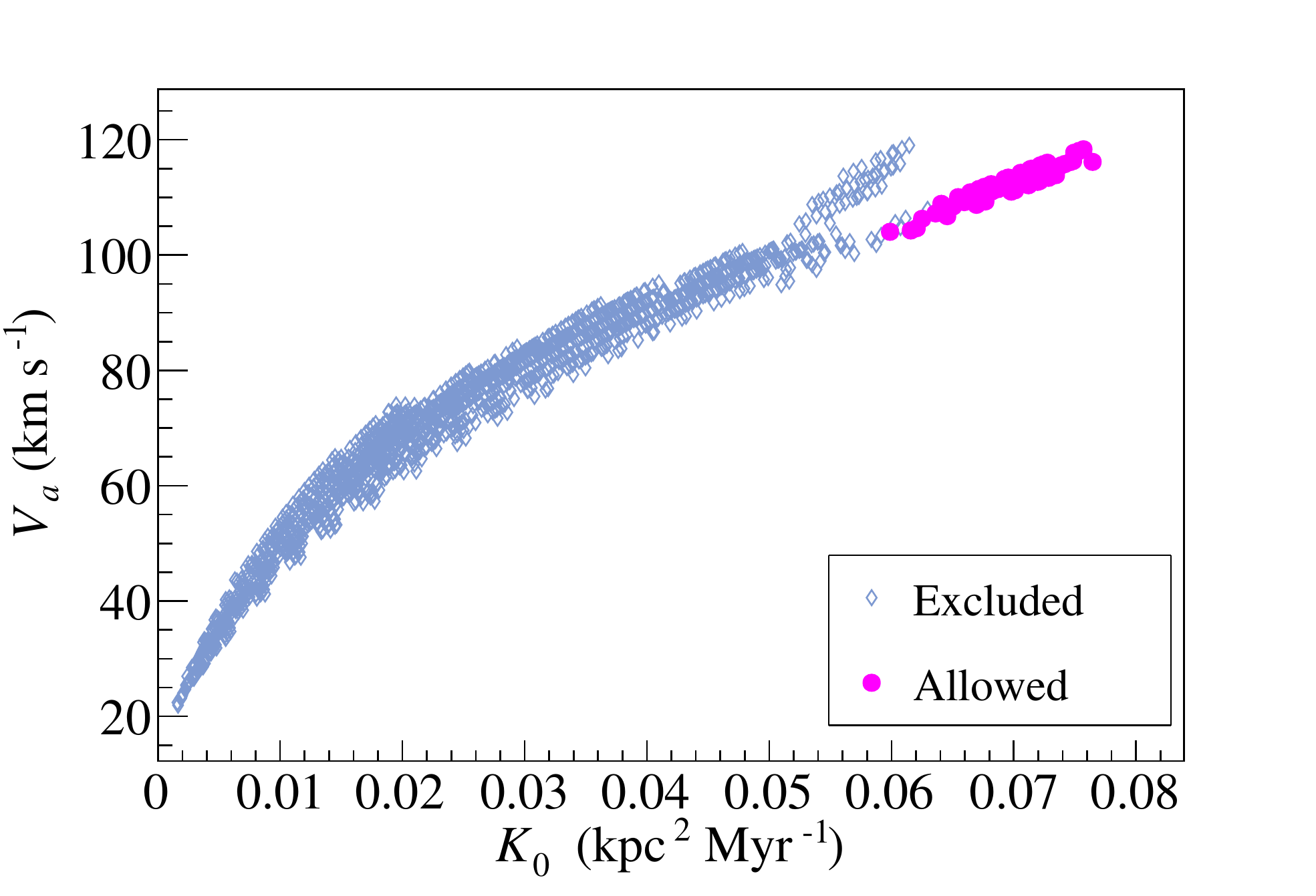}
\includegraphics[width=0.99\columnwidth]{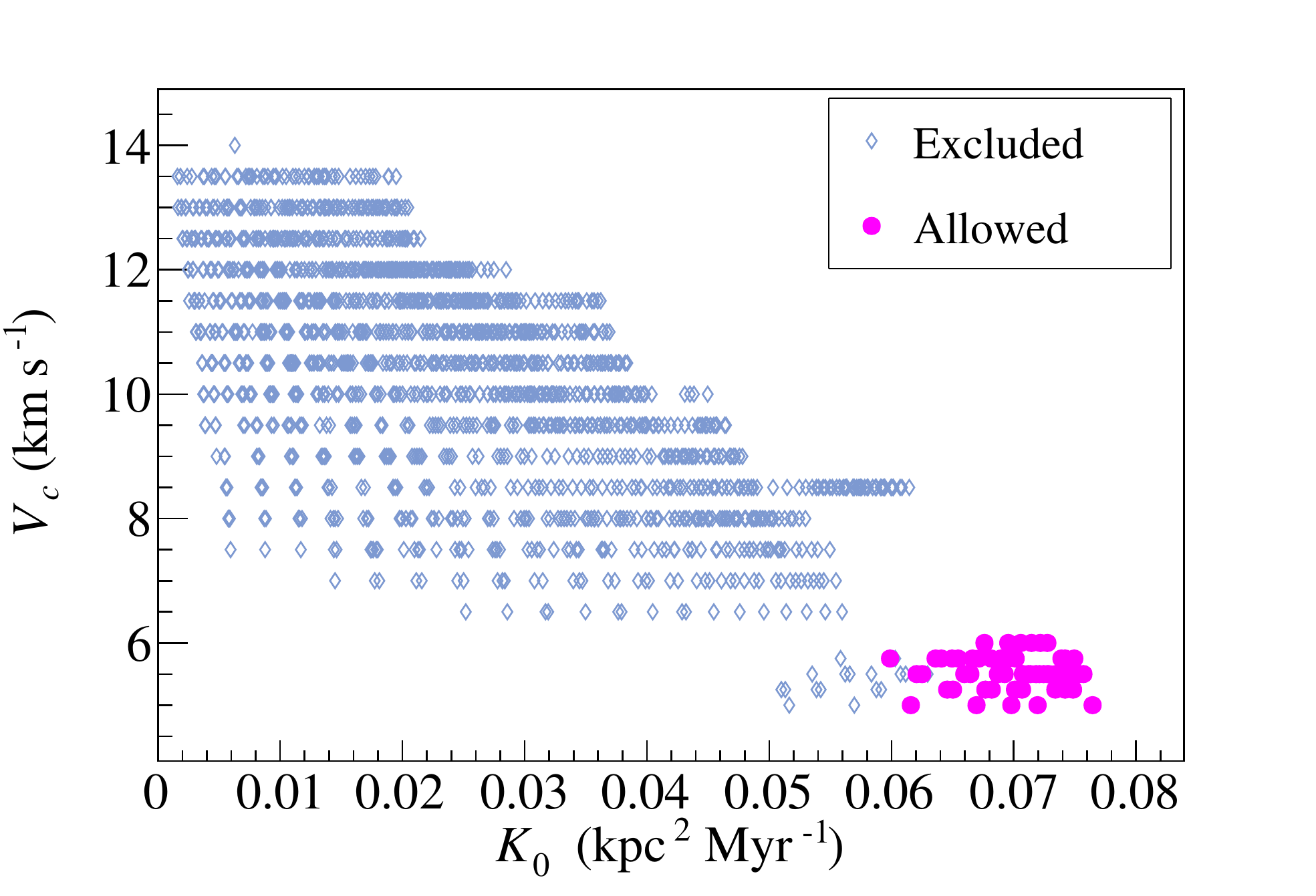}
\caption{Projection of the 1,623 propagation parameter sets selected by the B/C ratio\citepads{2001ApJ...555..585M} in the $K_0$ -- $L$, $K_0$ -- $\delta$, $K_0$ -- $\Va$, and $K_0$ -- $\Vc$ planes. The light blue diamonds show the propagation parameter values which are excluded, whereas the magenta circles denote the values which are allowed by this analysis.}
\label{fig:propagation_parameters}
\end{center}
\end{figure*}

An illustration of the selection method is presented in the left panel of Fig.~\ref{fig:ecremage}. In this figure, we display the \AMS positron flux and superimpose a colored band whose edges correspond to the envelope of the 1,623 predictions for the secondary positrons. The red colored region represents predictions that overshoot the data according to our definition and therefore contains the excluded models. On the other hand, the yellow colored region contains all allowed models. As an example, we display in dashed green a model that fulfills the Z-score constraint defined as $Z_{s} = \underset{E_i\,\in\,\text{data}}{\max} \,(Z_i) <3$, and in dashed red two models that do not respect it. The right panel of Fig.~\ref{fig:ecremage} illustrates the allowed propagation models that remain {\em after} the selection process: only 54 propagation sets out of 1,623 survive the criterion. Interestingly, one can see that the positron excess measured by \AMS seems to start already above 2 GeV, and not 10 GeV as often advocated. This will reveal itself very complicated to explain in terms of a single primary component. Selected models are those which minimize the secondary production over the whole energy range. Figure~\ref{fig:propagation_parameters} compares ranges of selected parameters with respect to their initial ones.
One can see that our new method enables us to drastically reduce the allowed parameter space with respect to former B/C analysis. Furthermore, we confirm the lifting of the degeneracy between $K_0$ and $L$, as one can see from the top-left panel of Fig.~\ref{fig:propagation_parameters}, as well as the high sensitivity to parameters that (mainly) control propagation at low energies, namely $\Va$ and $\Vc$. Pratically, common characteristics of these models are i) a large halo size $L$ (ranging from 8.5 to 15 kpc) together with relatively high $K_0$, typically $\geq 0.06$ kpc$^2$ Myr$^{-1}$, ii) a slope of the diffusion coefficient $\delta$ equal to 0.46, the minimal value allowed by the B/C analysis used in this study,  iii) small values of the convective wind $\Vc \leq 6$ km s$^{-1}$ accompanied by large values of the Alfv\`{e}n waves velocity $\Va \geq 100$ km s$^{-1}$. The fact that, in our analysis, $\delta$ is confined to the edge of the range indicates that even smaller values are likely to be favoured by positron data. This affirmation is indeed confirmed since, during the writing of this article, \AMS\ published the value of $\delta = 0.333 \pm 0.015$ from a power law fit of the high rigidity pure diffusive regime of their B/C data\citepads{PhysRevLett.117.231102}.
\vskip 0.1cm
These features can be readily understood. As reminded above, the secondary positron flux scales with $1/\sqrt{K_0}$. Hence, models with larger $K_0$ result in lower density of positrons at Earth compared to models with small diffusion coefficient. Given that secondary-to-primary ratios mostly constrain the $K_0/L$ ratio, selected models have a relatively high $L$, as well as a small value of $\delta$, the former beeing anti-correlated with $K_0$. Finally, values of the selected couples $\{\Va,\Vc\}$ minimize the bump at low energies and are therefore favored by the analysis.
Interestingly, in the recent literature, models with a large halo size have been suggested by other observables. Especially, the study of the antiproton-to-proton\citepads{2016PhRvL.117i1103A} and boron-to-carbon\citepads{2016PhRvL.117w1102A} ratios measured by \AMS point as well towards \MAX-like propagation model \citepads{2015JCAP...09..023G,2016arXiv160706093K}.  Radioactive species such as $^{10}{\rm Be}/^{9}{\rm Be}$ \citepads{2001AdSpR..27..717S,2010A&A...516A..66P} hint also at similar models although the dependence of this observable on the local density (local bubble) may bias the result. At other wavelengths (e.g. radio) \citepads{2013JCAP...03..036D} and in diffuse gamma ray analyses \citepads{2012PhRvD..85j9901A}, a high value of $L$ also seems to be preferred.  Even more recently, it has been shown that, as far as the stochastic injection of cosmic rays is concerned, the regularity of the proton spectrum could arise from a large magnetised halo size\citepads{2016arXiv161002010G}. Our results are in very good agreement with all these different observables, which therefore all underline the need for a primary positron component in order to explain data above a few GeV.
In the next section, we investigate the consequences of our updated propagation constraints on the hypothesis of dark matter annihilations as the source of this primary component.

%
%
%
%



\section{Dark matter interpretation of the A{\sc ms}-02 data}
\label{sec:DM_interpretation}


\subsection{Dark matter fitting procedure}

The most striking feature of the positron flux data is the high-energy gap with respect to the secondary prediction. Filling this gap with a dark matter component has been the concern of many studies, but semi-analytical methods were always restricted above 10 GeV (see for example\citetads{2015A&A...575A..67B}). Hereafter, we use the resolution method of Sect.~\ref{sec:pinching} to compute the positron flux following dark matter annihilation over all the energy range covered by \AMS data. Due to the important statistics of data below 10 GeV, constraints based only on the quality of the fit become more stringent.

Technically, we use the 54 propagation models selected in Sect.~\ref{sec:prop_param} to compute the theoretical prediction of the positron flux, which is the sum of a primary component coming from dark matter annihilation and the secondary component,
\beq
\Phi_{e^+}^{\rm{th}}=\Phi_{e^+}^{\rm{DM}}+\Phi_{e^+}^{\rm{II}}\,.
\eeq

We consider two different cases: DM particle annihilating into a general final state composed of quarks, leptons, and bosons, and the case of a leptophilic DM which annihilates into a combination of leptonic channels through a light mediator.

In a similar vein as\citetads{2015A&A...575A..67B}, we make no assumptions about the underlying DM model and consider the possibility that DM annihilates into a combination of channels, namely $b \bar{b}$, $W^+W^-$, $e^+e^-$, $\mu^+\mu^-$, and $\tau^+\tau^-$, with a branching ratio free to vary. The limited choice of these channels relies on the fact that they describe relatively well the various spectrum shape, and avoids introducing too many free parameters. For example, the $b\bar{b}$ channel typically describes the spectra of the different quark and gluon final states. To a certain extent positron spectra following Higgs decay are also similar to the $b\bar{b}$ case, since the Higgs decays dominantly into hadrons. Finally, the $W^+W^-$ channel is chosen to describe positron spectrum from gauge bosons decay. On the other hand, given the high dependence of the spectra on the lepton flavour, we allow non-universal lepton contributions. The DM annihilation spectra of all these channels are calculated using micrOMEGAs\_3.6\citepads{2011CoPhC.182..842B,2014CoPhC.185..960B}.

Concerning the case of a leptophilic DM, only three branching ratios are introduced as free parameters. They correspond to the three leptonic channels $(\phi\phi\rightarrow2e^+2e^-,\phi\phi\rightarrow2\mu^+2\mu^-, \phi\phi\rightarrow2\tau^+2\tau^-)$, where $\phi$ is a light scalar mediator. In this case the annihilation spectra are taken from the PPPC4DMID\citepads{2011JCAP...03..051C,2011JCAP...03..019C}.

For both cases, the DM component thus depends on the branching ratios, on the DM mass $m_\chi$, and on $\langle \sigma v\rangle$ the velocity averaged annihilation cross section, henceforth loosely dubbed "the cross section".

The search for the best fit to the positron data is led in the following way: for twenty DM masses logarithmically distributed in the range [100~GeV ; 1000~GeV], we perform a fit to the \AMS\ measurements of the positron flux using MINUIT. We determine the minimum value of the $\chi^2$ defined as
\beq
\chi^{2} = \sum_{i} \left\{ \frac{{\rm \Phi}^{\rm \, data}(E_i) - {\rm \Phi}^{\rm \, th}(E_i)}{\sigma^{\rm data}(E_i)} \right\}^{2} \, .
\label{eq:chi_2_def}
\eeq
In the case of the five annihilation channels, the parameter space is of dimension six: two corresponding to $m_\chi$ and $\langle \sigma v\rangle$, and four for the branching ratios $b_i$ given the constraint $\sum_i b_i = 1$. In the case of the leptophilic DM, the parameter space is of dimension four. To remain conservative, for each propagation model, we perform the fit seven times, varying the Fisk potential in the $3\sigma$ range [647~MV ; 830~MV] where 724 MV corresponds to the nominal value of the potential\citepads{2016A&A...591A..94G}.
In the following, we first discuss results for DM annihilation into the five channels $b \bar{b}$, $W^+W^-$, $e^+e^-$, $\mu^+\mu^-$, and $\tau^+\tau^-$, then for the leptophilic DM case.

\subsection{Results of the analysis}
\begin{figure*}[ht!]
\begin{center}
\includegraphics[width=0.99\columnwidth]{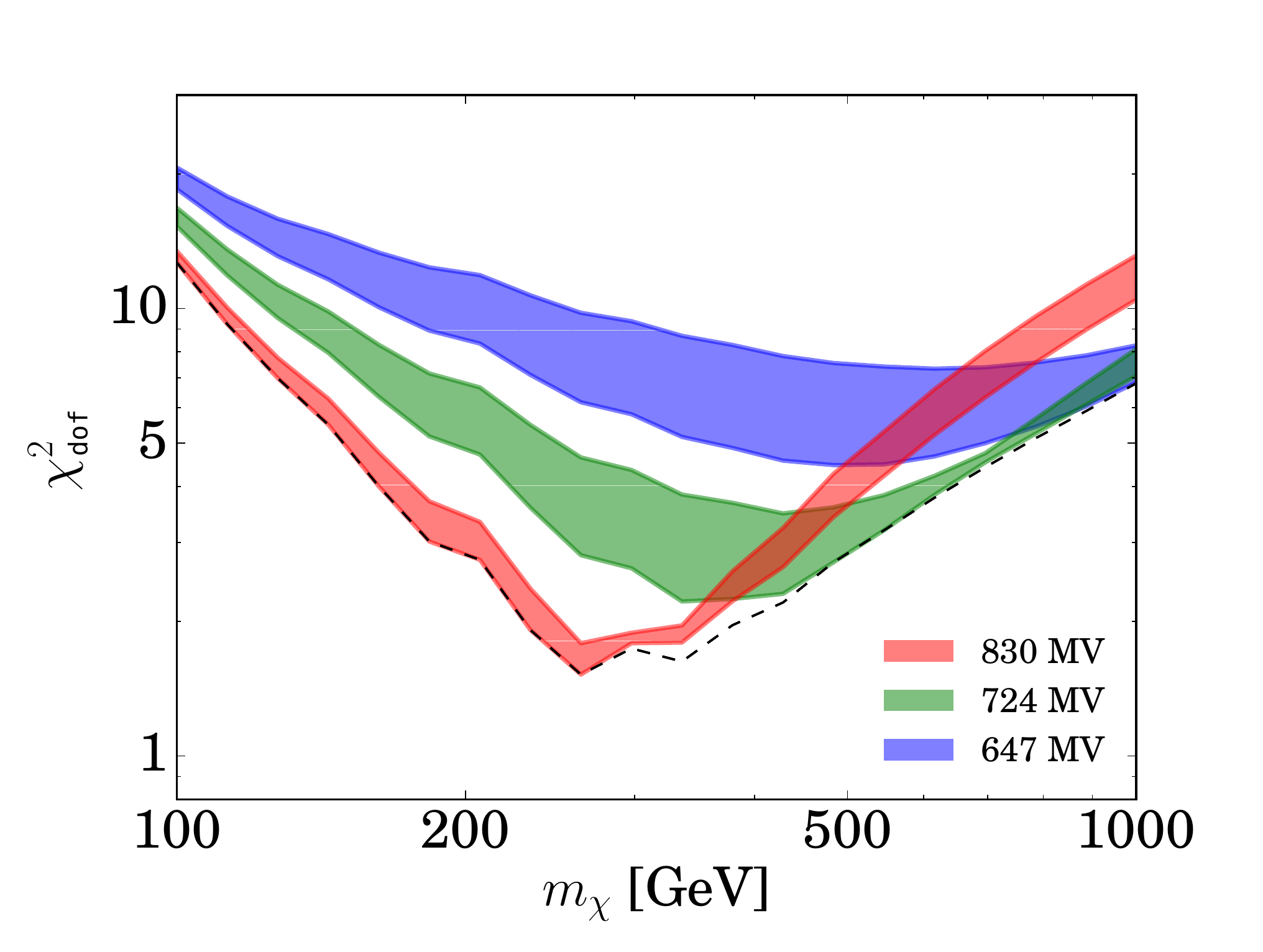}
\includegraphics[width=0.99\columnwidth]{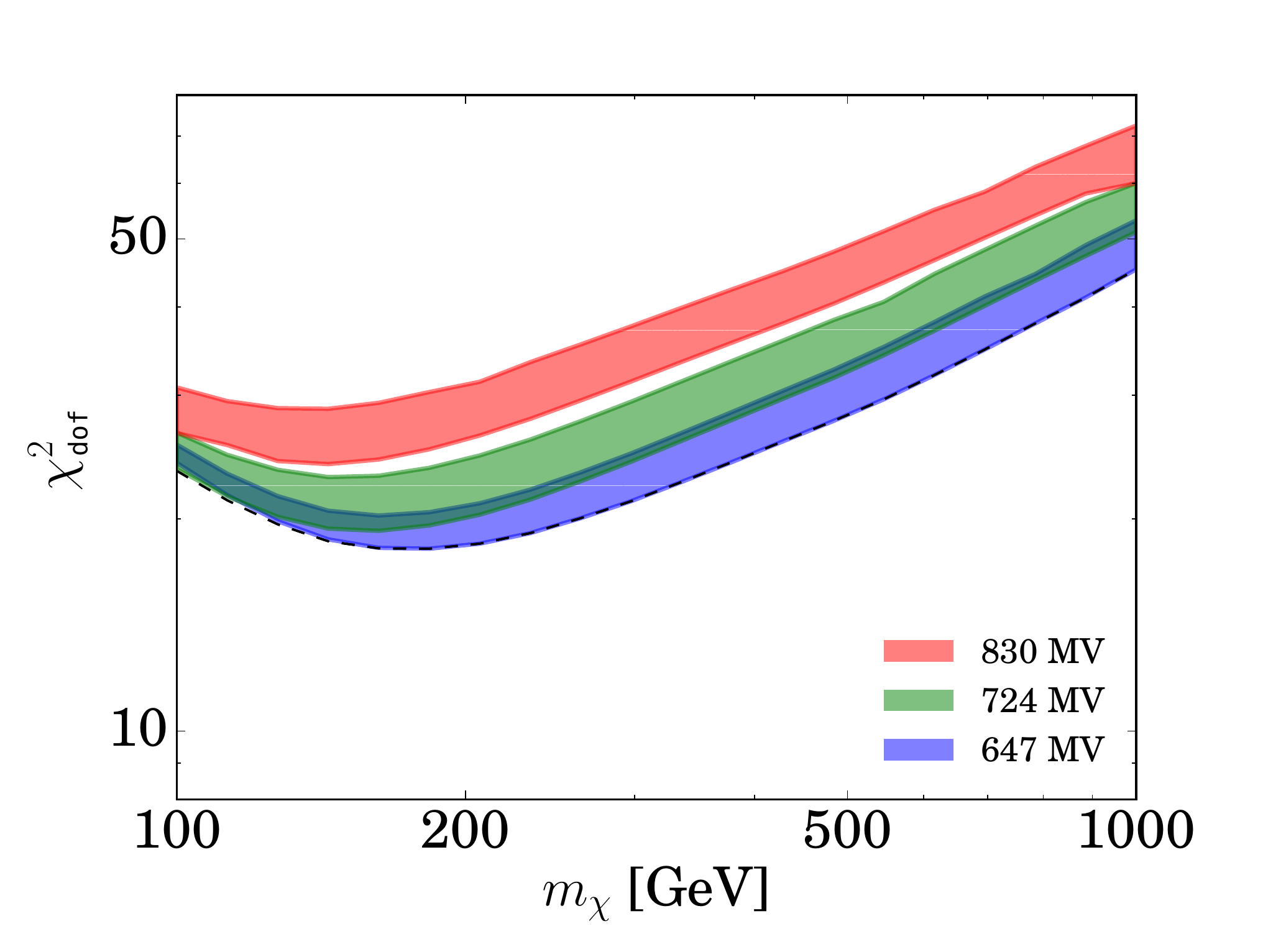}

\caption{Evolution of the $\chi^2_\mathrm{dof}$ as a function of the DM mass $m_\chi$ in the case of direct annihilation into standard model particles \textbf{(left panel)} and annihilation into four leptons through light mediators \textbf{(right panel)}. The results of the analysis are displayed using a Fisk potential of 830 MV, 724 MV, and 647 MV in red, green, and blue, respectively. The black dashed line represents the minimal $\chi^2_\mathrm{dof}$ among the seven Fisk potentials and 54 propagation models. The thickness of the colored band is obtained by scanning over the 54 propagation models.}
\label{fig:chi2_vs_mass}
\end{center}
\end{figure*}

We plot in Fig.~\ref{fig:chi2_vs_mass} the main result of our analysis, namely the evolution of the $\chi^2$ per degrees of freedom $\chi^2_\mathrm{dof}$, as a function of the DM mass $m_\chi$. The two plots correspond to DM annihilating into a fitted combination of $b \bar{b}$, $W^+W^-$, $e^+e^-$, $\mu^+\mu^-$, and $\tau^+\tau^-$ channels (left panel) and $\phi\phi\rightarrow2e^+2e^-$, $\phi\phi\rightarrow2\mu^+2\mu^-$, and $\phi\phi\rightarrow2\tau^+2\tau^-$ channels (right panel).  The results are displayed for different values of the Fisk potential (nominal value, and $\pm 3\sigma$). In the direct annihilation case, as one can see from Fig.~\ref{fig:chi2_vs_mass}, we find a a global best fit corresponding to a minimal $\chi^2_\mathrm{dof}=\chi^2/\mathrm{ndof}=100/66=1.5$. It is obtained for a DM mass of $m_\chi=264$~GeV annihilating into $b \bar{b}$, $e^+e^-$, and $\mu^+\mu^-$ with branching ratios of 0.92, 0.05, and 0.03 respectively (the branching ratios for the channels $W^+W^-$ and $\tau^+\tau^-$ are found to be zero). The associated annihilation cross section is $\sim$272 times larger than the thermal cross section. It means that a peculiar enhancement mechanism is required, as it has been found in many former studies.

Similarly, in the leptophilic case, we find a global best fit associated to a $\chi^2_{\rm dof}=1231/68=18$. It corresponds to a DM mass $m_\chi$ of $183$~GeV annihilating into $\phi\phi\rightarrow2e^+2e^-$ and $\phi\phi\rightarrow2\tau^+2\tau^-$ with respective branching ratios of 0.09 and 0.91. The branching ratio of the channel $\phi\phi\rightarrow2\mu^+2\mu^-$ is chosen as zero by the fit.

\begin{figure*}[ht!]
\begin{center}
\includegraphics[width=0.8\textwidth]{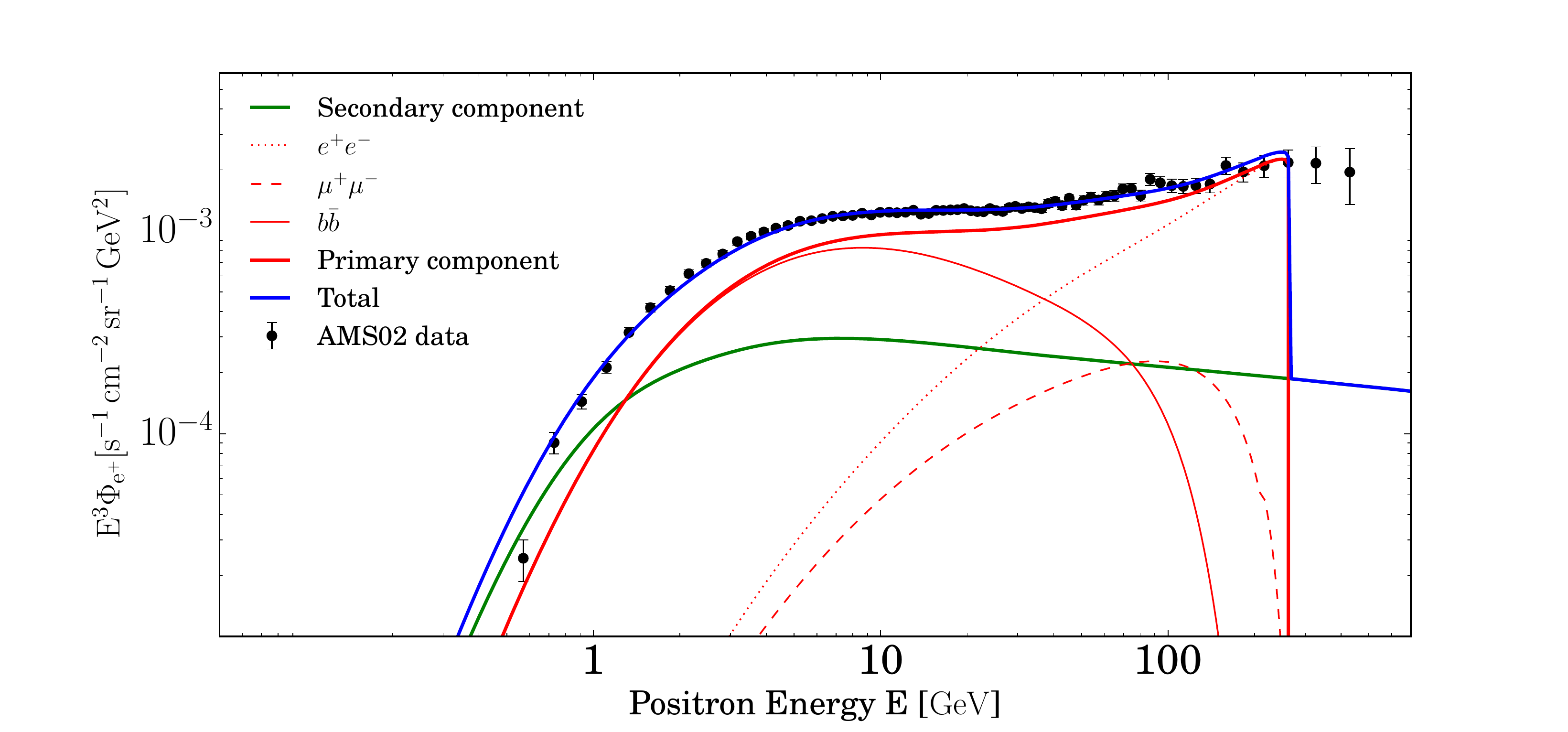}
\includegraphics[width=0.8\textwidth]{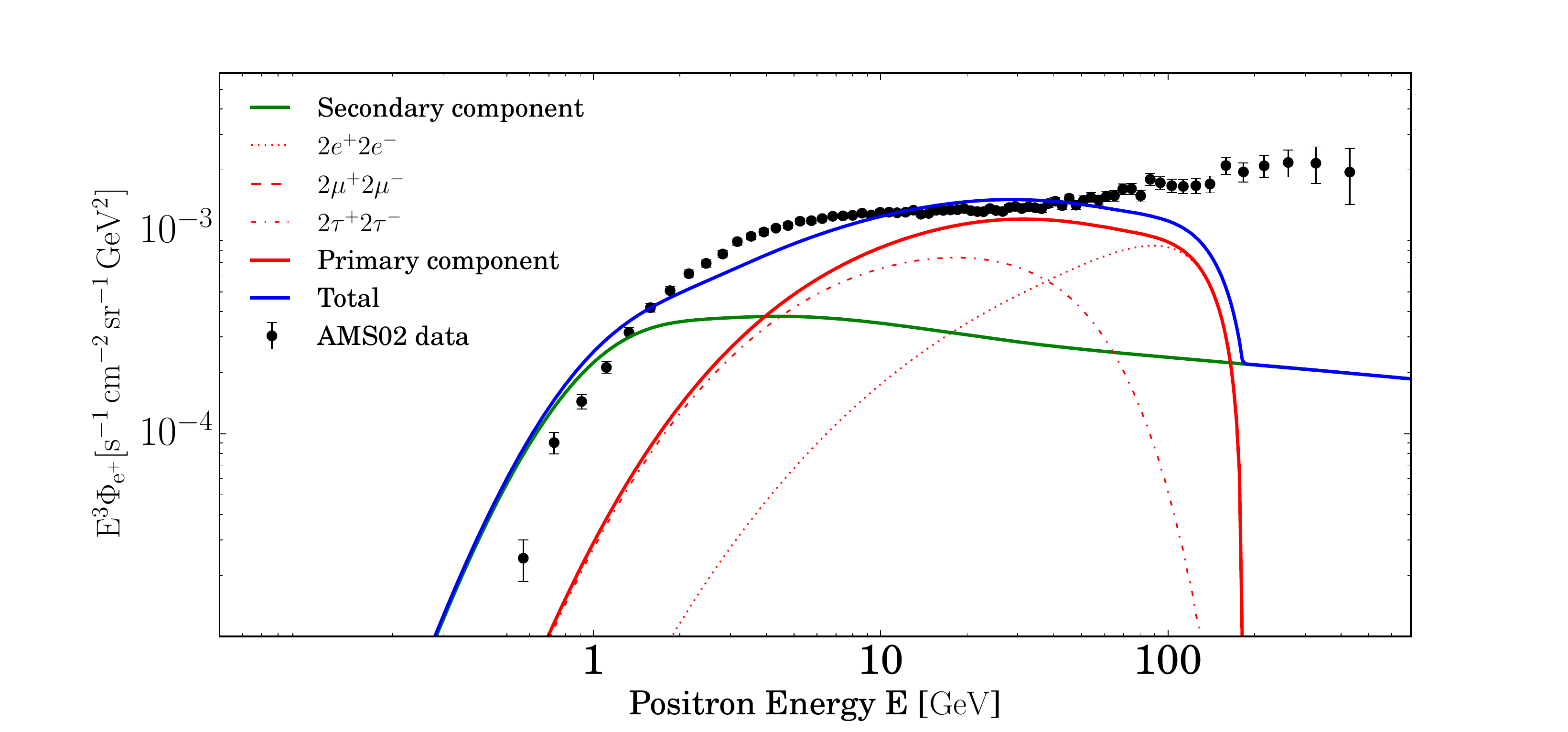}
\caption{Global best fit for the direct annihilation into standard model particles \textbf{(upper panel)} and annihilation into four leptons through light mediators \textbf{(lower panel)}. The DM (resp. secondary) component is displayed in thick red (resp. thick green) while the total flux, the sum of these two components, is shown in thick blue. In the case of direct annihilation, the electron, muon, and $b$ quark channels are displayed in dotted, dashed, and solid red line respectively. In the case of annihilations through light mediators, the electron and $\tau$ channels are displayed in dotted and dot-dashed red line respectively. The \AMS positron flux, including experimental uncertainties, is superimposed with black dots.}
\label{fig:best_fit_with_first_point}
\end{center}
\end{figure*}

Interestingly, values of the minimal $\chi^2_{\rm dof}$ are high, especially in the leptophilic case.
To understand results of the fitting procedure, we plot on Fig.~\ref{fig:best_fit_with_first_point} the theoretical positron fluxes obtained using the best fit models, together with the data.
In the direct annihilation case, one can note the remarkably good agrement of the fit with the data up to 300 GeV. However, the prediction is in discrepancy with the last two data points at two to four sigma. These two points (and marginally the first one) are responsible for the low quality of the fit yielding a $\chi^2_\mathrm{dof}=1.5$ or equivalently a $p$-value of 0.4\%. From left panel of Fig.~\ref{fig:chi2_vs_mass}, we observe that imposing the DM mass to be above 450 GeV in order to explain the last two points of the positron flux would yield an even poorer $\chi^2_\mathrm{dof}$, above 2. In the leptophilic case, the picture is even worse: no single part of the spectrum can be accurately described when one tries to fit the whole energy range. Thus, the resulting minimal $\chi^2_\mathrm{dof}$ is extremely bad.

Let us now discuss the evolution of $\chi^2_\mathrm{dof}$ with respect to the DM mass. First of all, we observe that, whatever is the solar modulation, the evolution of the $\chi^2_\mathrm{dof}$ is similar: with increasing DM mass, the $\chi^2_\mathrm{dof}$ first decreases, reaching a minimal values around a few hundred GeV, and then increases. Low DM masses cannot account for the high-energy part of the positron flux since no positrons with energy above the DM mass can be emitted. Thus, at first, the goodness of the fit is improving (i.e. the $\chi^2_\mathrm{dof}$ decreases) with the DM mass. Interestingly, above a peculiar DM mass, none of the channels can produce low-energy positrons in a sufficient amount to explain the low energy part of the data. Consequently the goodness of fit degrades, i.e. the $\chi^2_\mathrm{dof}$ increases. As a result, there is a ``middle ground'' at a peculiar mass (the value changes with annihilation channels and Fisk potential) which corresponds to the best possible attempt to fulfill similarly high- and low-energy constraint. Somehow, the flatness of the spectrum is such that it is not possible to accomodate it entirely with a single primary component.
We also note the drift of the best fit towards lower DM masses as the solar modulation increases. This is simply because the low energy part of the fluxes is more and more suppressed with an increasing Fisk potential. Hence, additional low-energy positrons are needed  (i.e. lighter DM) to fit the data. However increasing the Fisk potential is not necessary associated with an improving $\chi^2_\mathrm{dof}$: the actual shape of the annihilation spectrum matters, as it can be seen by comparing the upper and lower panels of Fig.~\ref{fig:best_fit_with_first_point}. Indeed, in the direct annihilation case increasing the Fisk potential tends to improve the fit, whereas in the case of annihilation through light mediators it worsens it.

In summary, we find challenging to interpret the excess in terms of pure DM annihilations, since our conservative analysis always leads to low-quality fits of the data. It is remarkable that the shape of the positron excess, with respect to the pure secondary prediction, cannot be captured by annihilations of a single DM species. This feature is due to: i) the high precision of the \AMS data; ii) the possibility to fit the whole data range thanks to our new semi-analytical resolution method. It is reasonable to believe that a fit above 10 GeV would not have had this issue. Similarly, we expect multi-component models, with e.g. one heavy and one light DM species to be able to fit the excess.  In the next section, we discuss how robust this conclusion is against a relaxation of our selection criterion of propagation parameters, as well as the inclusion of theoretical uncertainties in the modelling.

%

%
%



\section{Robustness of the results}
\label{sec:robustness}

In this section, we assess the robustness of the conclusions drawn above, under changes in the skimming method and source term modelisation. Since our selection criterion does not combine information of data points together but makes use of them separetely, we will investigate first whether a specific data point of the positron flux can be leading the constraints, eventually biasing the results. Indeed, one can see on Fig.~\ref{fig:ecremage} that the position of the first data point measured by \AMS is very low with respect to the expected trend from the predicted secondary positron flux. Secondly, we evaluate uncertainties of the secondary component source term in order to attest that they can be safely neglected in our analysis. These uncertainties come from the experimental measurement of the \AMS primary fluxes, as well as the choice of $p$ and He spallation cross section onto the ISM.

\begin{figure*}[ht!]
\begin{center}
\includegraphics[width=0.49\textwidth]{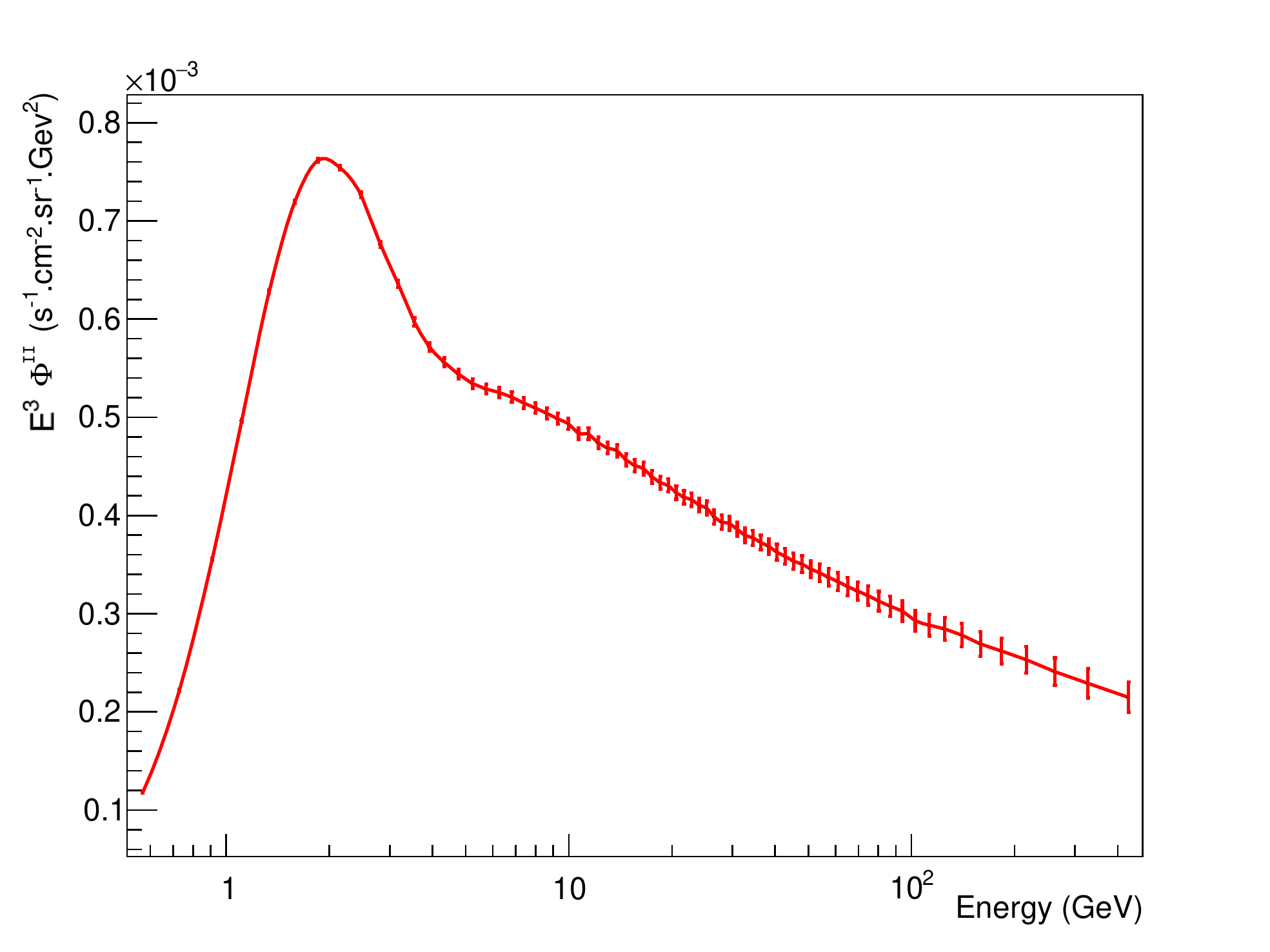}
\includegraphics[width=0.49\textwidth]{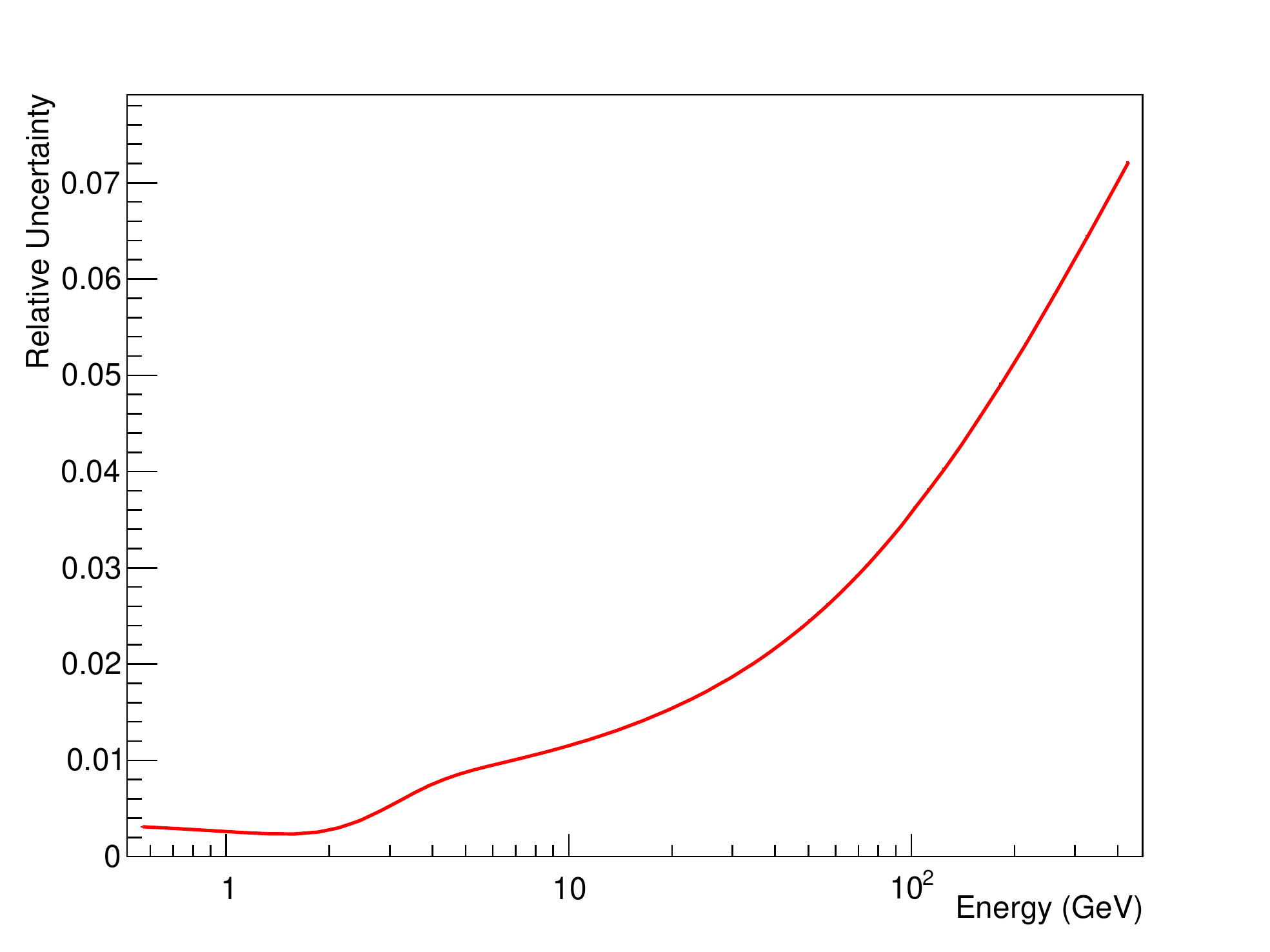}
\caption{{\bf Left panel:} secondary positron flux (multiplied by $E^3$) as a function of the positron energy. The error bars represent the uncertainty due to the experimental uncertainty on the proton and helium fluxes. {\bf Right panel:} relative uncertainty on the secondary positron flux, as a function of the positron energy.}
\label{fig:SecondaryUncertainty}
\end{center}
\end{figure*}

To check whether the first data point is more discriminating than the higher energy ones, we repeat the skimming method presented in Sect.~\ref{sec:prop_param} discarding this peculiar point from the analysis.
The comparison between the results of the analysis with and without the first point is reported in the first two lines of Table~\ref{tab:number_sets}. Not surprisingly, the number of selected models do increase. However, in a much striking way than expected, we notice that it increases more than twelve times. The parameter space counts now 623 allowed models. We conclude that, within our skimming method, the first point of the flux has indeed a very strong discriminating power. To check that it was a pecularity of the first data point, we repeat successively the skimming method discarding up to the three first data points. The results are reported in Table~\ref{tab:number_sets} and confirm the singularity of the first point: the number of allowed models never exceeds 692. Let us emphasize that even without its first point, the positron flux provides stringent constraints on propagation parameters: it enables to rule out two thirds of the parameter space allowed by former boron-over-carbon analysis.
To check the impact of a bigger parameter space on our DM analysis  (see Sect.~\ref{sec:DM_interpretation}), we repeat it with the 623 propagation models selected without the first point of the positron flux. In the case of DM annihilating directly into a combination of $b \bar{b}$, $W^+W^-$, $e^+e^-$, $\mu^+\mu^-$, and $\tau^+\tau^-$ channels, the $\chi^{2}_\mathrm{dof}$ of the best fit is now of $1.1$, which corresponds to a $p$-value of $26\%$. Such a value might indicate that DM annihilation can still explain the positron excess. However, the associated DM mass is $336$~GeV, causing a cut-off of the primary positron flux at this energy, not observed in \AMS data. Hence, with improving statistics in this last two bins, it is likely that the $\chi^{2}_\mathrm{dof}$ will quickly degrade. On the other hand, imposing the DM mass to be above the energy of the last data point increases the $\chi^{2}_\mathrm{dof}$ to a value above 2, synonym of a bad quality fit. In the hypothesis of leptophilic DM annihilating into $\phi\phi\rightarrow2e^+2e^-$ and $\phi\phi\rightarrow2\tau^+2\tau^-$ through a light mediator, the best fit has a $\chi^{2}_\mathrm{dof}$ bigger than 10. Thus, the conclusion remains unaltered.
%
\begin{table*}[ht!]
\centering
\begin{tabular}{c c c c c c c}
\hline
\hline
& Allowed & $\delta$ & $K_0$ & $L$ & $\Vc$ & $\Va$ \\
& propagation models &  & (kpc$^2$ Myr$^{-1}$) & (kpc) & (km s$^{-1}$) & (km s$^{-1}$) \\
\hline
\CellTop
All data points& 54 & 0.46 & 0.0599 -- 0.0764 & 8.5 -- 15 & 5 -- 6 & 104.0 -- 118.3 \\
First point excluded& 623 & 0.46 -- 0.7 & 0.0240 -- 0.0764 & 4.5 -- 15 & 5 -- 12 & 70.9 -- 119.0 \\
First two points excluded& 623 & 0.46 -- 0.7 & 0.0240 -- 0.0764 & 4.5 -- 15 & 5 -- 12 & 70.9 -- 119.0 \\
First three points excluded& 692 & 0.46 -- 0.7 & 0.0215 -- 0.0764 & 4 -- 15 & 5 -- 12 & 70.4 -- 119.0 \\
\hline
\end{tabular}
\vspace{0.5cm}
\caption{Number of propagation models allowed after the analysis of Sect.~\ref{sec:prop_param}, and associated parameter ranges. We present results of the skimming method discarding successively up to the three first data points. \label{tab:number_sets}}
\end{table*}
%
We now turn to assessing the impact of uncertainties associated to the source term of the secondary component on our conclusions.
A key ingredient of the secondary positron prediction is an accurate measurement of the flux of their progenitors, mainly proton and helium nuclei.  In Sect.~\ref{sec:astrobkg}, we gave the parameterisation used to describe these fluxes, as well as the best-fit value of the parameters. Given the finite precision of \AMS measurements, uncertainties in the determination of these parameters can affect our secondary positron prediction. To estimate the uncertainty associated to the fitting procedure, we developed an original method that takes into account both systematic and statistical uncertainties of the measured primary fluxes. We proceed in the following way: we first generate mock data of the primary fluxes within their total uncertainties, fit them with our parameterisation and compute a new secondary positron flux. Repeating this process 10,000 times allows us to determine the distribution of the secondary positron flux in each energy bin. The mock data for the primary fluxes are generated according to the following strategy: for each data point a new random value is computed as $\bar{\Phi}^{\rm data}(E_i) + \delta\Phi^{\rm stat}(E_i) +\delta\Phi^{\rm syst}(E_i)$, where $\bar{\Phi}^{\rm data}$ is the mean value of the flux in the energy bin $E_i$,  $\delta\Phi^{\rm stat}$ is drawn from a Gaussian distribution with standard deviation $\sigma^{\rm stat}(E_i)$ and $\delta\Phi^{\rm syst}$ is drawn from a uniform distribution of size 2$\sigma^{\rm syst}(E_i)$. These two uncertainties $\sigma$ are provided by the \AMS collaboration in\citetads{2015PhRvL.114q1103A} and\citetads{2015PhRvL.115u1101A}. Results are displayed in Fig.~\ref{fig:SecondaryUncertainty}. On the left panel is shown the distribution of our prediction in each energy bin, compared to the fiducial value calculated with the \MED\ propagation model. The relative uncertainty displayed on the right panel is found to increase with the energy, with a maximum of $7\%$ at 500 GeV.
The experimental uncertainties of the positron flux are respectively of $6\%$ and $30\%$, much larger than the theoretical uncertainty yielded by the primary fluxes. We thus conclude that the precision in the measurement of the primary fluxes is sufficiently small not to alter our analysis.\\
A second major ingredient entering the source term for secondary positrons are the cross sections adopted for the $p$ and He interaction with the ISM. In our studies, we used proton-proton cross section from\citetads{2006ApJ...647..692K}. We recall that any other nucleus-nucleus cross section can be obtained by rescaling this one with an empirical factor, which we took from\citetads{2007NIMPB.254..187N}. The choice of proton-proton cross section from\citetads{2006ApJ...647..692K} is motivated by the fact that, at low energy, this model produces {\em less} positrons than the commonly used\citetads{1998ApJ...493..694M}, which includes the parameterisation of the Lorentz invariant obtained by\citetads{1983JPhG....9.1289T} and\citetads{1977PhRvD..15..820B}.  We therefore adopt a strategy similar to our treatment of solar modulation, which minimizes as much as possible the positron flux below 10 GeV by using a very high Fisk potential, on the edge of current allowed values. Although there is an uncertainty associated to the cross section and solar modulation modelling, our choices lead to conservative results and thus robust conclusions.

%
%



\section{Conclusion}
\label{sec:conclusion}

Two years ago, the \AMS collaboration released the most precise measurement of the positron flux in the energy range 0.5 to 500~GeV, confirming the high-energy excess with respect to pure secondary predictions. Until now, most of the studies trying to explain this excess in terms of DM annihilations restricted themselves to energies above 10 GeV by prejudice and to simplify computations. Indeed, below this energy, several  mechanisms taking place in the halo, namely diffusive reacceleration and convection, as well as energy losses in the disc, make the resolution of the propagation equation much more involved. However, a consistent model should be able to explain the positron flux over the entire energy range covered by the \AMS\ data.

We have therefore reinvestigated the problem of the positron anomaly with a new semi-analytical resolution method enabling us to take into account transport processes so far neglected but important below a few GeV. The key idea is to pinch energy losses occurring in the whole magnetic halo, namely inverse Compton and synchrotron, inside the Galactic disc. The corresponding energy loss rate is artificially enhanced by a so-called pinching factor which is calculated for each energy. This allows us to solve the CR propagation equation using a Bessel expansion and the Cranck-Nicholson scheme.

With this procedure, we recover the correct high-energy positron flux at the per mille level and extend the computation to low energies, at the main advantage of a very fast resolution technique compared to fully numerical methods. We have re-evaluated both primary and secondary components of the positron flux, finding as expected that they are significantly affected at low energies by the incorporation of the so far neglected CR transport mechanisms.
Surprisingly, we also find that modifications are still substantial at a few tens of GeV, depending on the CR propagation parameters. As pointed out in\citetads{2009A&A...501..821D} and\citetads{2014PhRvD..90h1301L}, we confirm that the secondary prediction of the positron flux at low energies can be in large excess compared to \AMS data, even for propagation models compatible with the B/C analysis from\citetads{2001ApJ...555..585M}.

Motivated by this result, we have carried out a scan over the CR propagation parameters of\citetads{2001ApJ...555..585M}, applying an original skimming method which leads to severe constraints on the propagation parameters: out of the 1,623 models, only 54 survive the procedure. In particular, the benchmark {\MIN} and {\MED} configurations are excluded. On the other hand, {\MAX}-like propagation models, i.e. with large \{$K_0$,$L$\} and small $\delta$, are favored by the data. Those models are very close to the best configuration found in\citetads{2015JCAP...10..034K} by fitting the preliminary B/C data of \AMS. This needs to be confirmed with the newest B/C published recently by \AMS \citepads{PhysRevLett.117.231102}. We do not expect major changes in our conclusions.

To overcome the difficulty arising from solar modulation, we have made use of a very high Fisk potential, $3\sigma$ above the mean value obtained by\citetads{2016A&A...591A..94G}. This choice minimizes the flux at low energies and makes our skimming procedure of the CR parameter space conservative, leading us to keep models that should be disregarded.

In a similar way, we have made use of the $p$ and He spallation cross sections from\citetads{2006ApJ...647..692K} since they lead to the lowest amount of positrons. Furthermore, we have checked that uncertainties in the measurements of the $p$ and He fluxes do not alter our result. Finally, given that our skimming method makes use of information from data points separately, we have investigated whether a specific data point could drive the constraints. We found that this is indeed the case: the first data point has a much higher discriminating power than the others. Discarding it from the analysis, our skimming method selected 623 models, which still corresponds to a rejection of about two thirds of the parameter space. We can therefore conclude that the positron flux is a very useful and independent probe of CR propagation, to be used in synergy with other tracers such as the B/C ratio. Our results also indicate that the positron excess is already present at GeV energies, typically starting above 2 GeV.

We have finally re-investigated the explanation of the positron anomaly in terms of annihilations of a single DM species, in the WIMP framework, over the whole energy range of the \AMS data. We have performed a scan over WIMP mass and explored the possibility of: i) direct annihilation into a combination of channels; ii) leptophilic DM annihilating into four leptons through a light mediator. For a given WIMP mass and propagation model (selected by our skimming method), we have obtained the best-fit values of the annihilation cross section and branching ratios. Our most striking result is that no good fit to the data is obtained for both cases i) and ii). Indeed, in case i), the best fit is found for a mass of 264~GeV that does not allow to reproduce the highest-energy data points. Hence, the associated $p$-value is as low as 0.4\%. On the other hand, requiring the DM mass to be larger than 500~GeV yields $\chi^{2}_{\rm dof}>3$, since the low-energy part of the data cannot be consistently accommodated. Case ii) turns out to be even worse, the best-fit $\chi^{2}_{\rm dof}$ beeing as high as 15. We have checked the robustness of our conclusions against a few possible loopholes.

\vskip 0.1cm
We are thus led to the conclusion that annihilations of a single DM species should be disregarded as the sole origin of the positron excess, on the basis of the positron data themselves, irrespective of other observables such as the antiproton flux or CMB anisotropies. It is likely that more ad-hoc multi-species models, with e.g. one heavy and one light DM particle, will be able to accommodate the excess, although a strong statement would require a dedicated study.
It is probable that such an analysis with a unique pulsar as the source of the anomaly would lead to similar conclusions, requiring in the future more realistic multi-component studies.



\begin{acknowledgements}
We would like to thank Pasquale D. Serpico, Richard Taillet and Eric Ragoucy for enlightening discussions in the first stages of this work.
Part of this work was supported by the French \emph{Institut universitaire de France}, by the French
\emph{Agence Nationale de la Recherche} under contract 12-BS05-0006 DMAstroLHC, by the \emph{Investissements d'avenir}, Labex ENIGMASS, by the CNES, France, and by the PRC (\emph{Projet de Recherche Conjoint}) CNRS-FAPESP.
MV and EFB are grateful to the S\~ao Paulo Research Foundation (FAPESP) for the support received through grants no. 2014/19149-7, 2014/50747-8 and 2015/13533-2.
\end{acknowledgements}


\newpage
\bibliographystyle{aa}
\bibliography{CRAC_3_biblio}

\begin{thebibliography}{56}
\expandafter\ifx\csname natexlab\endcsname\relax\def\natexlab#1{#1}\fi

\bibitem[{{Accardo} {et~al.}(2014){Accardo}, {Aguilar}, {Aisa}, {Alvino},
  {Ambrosi}, {Andeen}, {Arruda}, {Attig}, {Azzarello}, {Bachlechner}, \&
  et~al.}]{2014PhRvL.113l1101A}
{Accardo}, L., {Aguilar}, M., {Aisa}, D., {et~al.} 2014, Physical Review
  Letters, 113, 121101

\bibitem[{{Ackermann} {et~al.}(2012){Ackermann}, {Ajello}, {Albert}, {Baldini},
  {Ballet}, {Barbiellini}, {Bastieri}, {Bechtol}, {Bellazzini}, {Bloom},
  {Bonamente}, {Borgland}, {Brandt}, {Bregeon}, {Brigida}, {Bruel}, {Buehler},
  {Buson}, {Caliandro}, {Cameron}, {Caraveo}, {Cecchi}, {Charles}, {Chekhtman},
  {Chiang}, {Ciprini}, {Claus}, {Cohen-Tanugi}, {Conrad}, {Cuoco}, {Cutini},
  {D'Ammando}, {de Palma}, {Dermer}, {Digel}, {do Couto e Silva}, {Drell},
  {Drlica-Wagner}, {Dubois}, {Favuzzi}, {Fegan}, {Ferrara}, {Fortin},
  {Fukazawa}, {Fusco}, {Gargano}, {Gasparrini}, {Germani}, {Giglietto},
  {Giroletti}, {Glanzman}, {Godfrey}, {Gomez-Vargas}, {Gr{\'e}goire},
  {Grenier}, {Grove}, {Guiriec}, {Gustafsson}, {Hadasch}, {Hayashida},
  {Hayashi}, {Hou}, {Hughes}, {J{\'o}hannesson}, {Johnson}, {Kamae},
  {Kn{\"o}dlseder}, {Kuss}, {Lande}, {Latronico}, {Lemoine-Goumard}, {Linden},
  {Lionetto}, {Llena Garde}, {Longo}, {Loparco}, {Lovellette}, {Lubrano},
  {Mazziotta}, {McEnery}, {Mitthumsiri}, {Mizuno}, {Monte}, {Monzani},
  {Morselli}, {Moskalenko}, {Murgia}, {Naumann-Godo}, {Norris}, {Nuss},
  {Ohsugi}, {Okumura}, {Orienti}, {Orlando}, {Ormes}, {Paneque}, {Panetta},
  {Parent}, {Pavlidou}, {Pesce-Rollins}, {Pierbattista}, {Piron}, {Pivato},
  {Rain{\`o}}, {Rando}, {Reimer}, {Reimer}, {Roth}, {Sbarra}, {Schmitt},
  {Sgr{\`o}}, {Siegal-Gaskins}, {Siskind}, {Spandre}, {Spinelli}, {Strong},
  {Suson}, {Takahashi}, {Tanaka}, {Thayer}, {Tibaldo}, {Tinivella}, {Torres},
  {Tosti}, {Troja}, {Usher}, {Vandenbroucke}, {Vasileiou}, {Vianello},
  {Vitale}, {Waite}, {Winer}, {Wood}, {Wood}, {Yang}, {Zimmer}, \&
  {Komatsu}}]{2012PhRvD..85j9901A}
{Ackermann}, M., {Ajello}, M., {Albert}, A., {et~al.} 2012, \prd, 85, 109901

\bibitem[{{Adriani} {et~al.}(2009){Adriani}, {Barbarino}, {Bazilevskaya},
  {Bellotti}, {Boezio}, {Bogomolov}, {Bonechi}, {Bongi}, {Bonvicini}, {Bottai},
  {Bruno}, {Cafagna}, {Campana}, {Carlson}, {Casolino}, {Castellini}, {de
  Pascale}, {de Rosa}, {de Simone}, {di Felice}, {Galper}, {Grishantseva},
  {Hofverberg}, {Koldashov}, {Krutkov}, {Kvashnin}, {Leonov}, {Malvezzi},
  {Marcelli}, {Menn}, {Mikhailov}, {Mocchiutti}, {Orsi}, {Osteria}, {Papini},
  {Pearce}, {Picozza}, {Ricci}, {Ricciarini}, {Simon}, {Sparvoli},
  {Spillantini}, {Stozhkov}, {Vacchi}, {Vannuccini}, {Vasilyev}, {Voronov},
  {Yurkin}, {Zampa}, {Zampa}, \& {Zverev}}]{2009Natur.458..607A}
{Adriani}, O., {Barbarino}, G.~C., {Bazilevskaya}, G.~A., {et~al.} 2009, \nat,
  458, 607

\bibitem[{{Aguilar} {et~al.}(2015{\natexlab{a}}){Aguilar}, {Aisa}, {Alpat},
  {Alvino}, {Ambrosi}, {Andeen}, {Arruda}, {Attig}, {Azzarello}, {Bachlechner},
  \& et~al.}]{2015PhRvL.115u1101A}
{Aguilar}, M., {Aisa}, D., {Alpat}, B., {et~al.} 2015{\natexlab{a}}, Physical
  Review Letters, 115, 211101

\bibitem[{{Aguilar} {et~al.}(2015{\natexlab{b}}){Aguilar}, {Aisa}, {Alpat},
  {Alvino}, {Ambrosi}, {Andeen}, {Arruda}, {Attig}, {Azzarello}, {Bachlechner},
  \& et~al.}]{2015PhRvL.114q1103A}
{Aguilar}, M., {Aisa}, D., {Alpat}, B., {et~al.} 2015{\natexlab{b}}, Physical
  Review Letters, 114, 171103

\bibitem[{{Aguilar} {et~al.}(2014){Aguilar}, {Aisa}, {Alvino}, {Ambrosi},
  {Andeen}, {Arruda}, {Attig}, {Azzarello}, {Bachlechner}, {Barao}, \&
  et~al.}]{2014PhRvL.113l1102A}
{Aguilar}, M., {Aisa}, D., {Alvino}, A., {et~al.} 2014, Physical Review
  Letters, 113, 121102

\bibitem[{{Aguilar} {et~al.}(2013){Aguilar}, {Alberti}, {Alpat}, {Alvino},
  {Ambrosi}, {Andeen}, {Anderhub}, {Arruda}, {Azzarello}, {Bachlechner}, \&
  et~al.}]{2013PhRvL.110n1102A}
{Aguilar}, M., {Alberti}, G., {Alpat}, B., {et~al.} 2013, Physical Review
  Letters, 110, 141102

\bibitem[{{Aguilar} {et~al.}(2016){Aguilar}, {Ali Cavasonza}, {Alpat},
  {Ambrosi}, {Arruda}, {Attig}, {Aupetit}, {Azzarello}, {Bachlechner}, {Barao},
  \& et~al.}]{2016PhRvL.117i1103A}
{Aguilar}, M., {Ali Cavasonza}, L., {Alpat}, B., {et~al.} 2016, Physical Review
  Letters, 117, 091103

\bibitem[{Aguilar {et~al.}(2016)Aguilar, Ali~Cavasonza, Ambrosi, Arruda, Attig,
  Aupetit, Azzarello, Bachlechner, Barao, Barrau, Barrin, Bartoloni, Basara,
  Ba\ifmmode \mbox{\c{s}}\else \c{s}\fi{}e\ifmmode \breve{g}\else
  \u{g}\fi{}mez-du Pree, Battarbee, Battiston, Becker, Behlmann, Beischer,
  Berdugo, Bertucci, Bindel, Bindi, Boella, de~Boer, Bollweg, Bonnivard,
  Borgia, Boschini, Bourquin, Bueno, Burger, Cadoux, Cai, Capell, Caroff,
  Casaus, Castellini, Cervelli, Chae, Chang, Chen, Chen, Chen, Cheng, Chou,
  Choumilov, Choutko, Chung, Clark, Clavero, Coignet, Consolandi, Contin,
  Corti, Creus, Crispoltoni, Cui, Dai, Delgado, Della~Torre, Demakov,
  Demirk\"oz, Derome, Di~Falco, Dimiccoli, D\'{\i}az, von Doetinchem, Dong,
  Donnini, Duranti, D'Urso, Egorov, Eline, Eronen, Feng, Fiandrini, Finch,
  Fisher, Formato, Galaktionov, Gallucci, Garc\'{\i}a, Garc\'{\i}a-L\'opez,
  Gargiulo, Gast, Gebauer, Gervasi, Ghelfi, Giovacchini, Goglov, G\'omez-Coral,
  Gong, Goy, Grabski, Grandi, Graziani, Guo, Haino, Han, He, Heil, Hoffman,
  Hsieh, Huang, Huang, Huh, Incagli, Ionica, Jang, Jinchi, Kang, Kanishev, Kim,
  Kim, Kirn, Konak, Kounina, Kounine, Koutsenko, Krafczyk, La~Vacca, Laudi,
  Laurenti, Lazzizzera, Lebedev, Lee, Lee, Leluc, Li, Li, Li, Li, Li, Li, Li,
  Li, Li, Lim, Lin, Lipari, Lippert, Liu, Liu, Lordello, Lu, Lu, Luebelsmeyer,
  Luo, Luo, Lv, Machate, Majka, Ma\~n\'a, Mar\'{\i}n, Martin, Mart\'{\i}nez,
  Masi, Maurin, Menchaca-Rocha, Meng, Mikuni, Mo, Morescalchi, Mott, Nelson,
  Ni, Nikonov, Nozzoli, Oliva, Orcinha, Palmonari, Palomares, Paniccia,
  Pauluzzi, Pensotti, Pereira, Picot-Clemente, Pilo, Pizzolotto, Plyaskin,
  Pohl, Poireau, Putze, Quadrani, Qi, Qin, Qu, R\"aih\"a, Rancoita, Rapin,
  Ricol, Rosier-Lees, Rozhkov, Rozza, Sagdeev, Sandweiss, Saouter, Schael,
  Schmidt, Schulz~von Dratzig, Schwering, Seo, Shan, Shi, Siedenburg, Son,
  Song, Sun, Tacconi, Tang, Tang, Tao, Tescaro, Ting, Ting, Tomassetti, Torsti,
  T\"urko\ifmmode~\breve{g}\else \u{g}\fi{}lu, Urban, Vagelli, Valente,
  Vannini, Valtonen, V\'azquez~Acosta, Vecchi, Velasco, Vialle, Vitale,
  Vitillo, Wang, Wang, Wang, Wang, Wang, Wang, Wei, Weng, Whitman,
  Wienkenh\"over, Wu, Wu, Xia, Xiong, Xu, Yan, Yang, Yang, Yang, Yi, Yu, Yu,
  Zeissler, Zhang, Zhang, Zhang, Zhang, Zhang, Zhang, Zheng, Zhu, Zhuang,
  Zhukov, Zichichi, Zimmermann, \& Zuccon}]{PhysRevLett.117.231102}
Aguilar, M., Ali~Cavasonza, L., Ambrosi, G., {et~al.} 2016, Phys. Rev. Lett.,
  117, 231102

\bibitem[{{Aguilar} {et~al.}(2016){Aguilar}, {Ali Cavasonza}, {Ambrosi},
  {Arruda}, {Attig}, {Aupetit}, {Azzarello}, {Bachlechner}, {Barao}, {Barrau},
  \& et~al.}]{2016PhRvL.117w1102A}
{Aguilar}, M., {Ali Cavasonza}, L., {Ambrosi}, G., {et~al.} 2016, Physical
  Review Letters, 117, 231102

\bibitem[{{Badhwar} {et~al.}(1977){Badhwar}, {Golden}, \&
  {Stephens}}]{1977PhRvD..15..820B}
{Badhwar}, G.~D., {Golden}, R.~L., \& {Stephens}, S.~A. 1977, \prd, 15, 820

\bibitem[{{Baltz} \& {Edsj{\"o}}(1999)}]{1999PhRvD..59b3511B}
{Baltz}, E.~A. \& {Edsj{\"o}}, J. 1999, \prd, 59, 023511

\bibitem[{{Barwick} {et~al.}(1997){Barwick}, {Beatty}, {Bhattacharyya},
  {Bower}, {Chaput}, {Coutu}, {de Nolfo}, {Knapp}, {Lowder}, {McKee},
  {Mueller}, {Musser}, {Nutter}, {Schneider}, {Swordy}, {Tarle}, {Tomasch},
  {Torbet}, \& {HEAT Collaboration}}]{1997ApJ...482L.191B}
{Barwick}, S.~W., {Beatty}, J.~J., {Bhattacharyya}, A., {et~al.} 1997, \apjl,
  482, L191

\bibitem[{{Beatty} {et~al.}(2004){Beatty}, {Bhattacharyya}, {Bower}, {Coutu},
  {Duvernois}, {McKee}, {Minnick}, {M{\"u}ller}, {Musser}, {Nutter},
  {Labrador}, {Schubnell}, {Swordy}, {Tarl{\'e}}, \&
  {Tomasch}}]{2004PhRvL..93x1102B}
{Beatty}, J.~J., {Bhattacharyya}, A., {Bower}, C., {et~al.} 2004, Physical
  Review Letters, 93, 241102

\bibitem[{{B{\'e}langer} {et~al.}(2011){B{\'e}langer}, {Boudjema}, {Brun},
  {Pukhov}, {Rosier-Lees}, {Salati}, \& {Semenov}}]{2011CoPhC.182..842B}
{B{\'e}langer}, G., {Boudjema}, F., {Brun}, P., {et~al.} 2011, Computer Physics
  Communications, 182, 842

\bibitem[{{B{\'e}langer} {et~al.}(2014){B{\'e}langer}, {Boudjema}, {Pukhov}, \&
  {Semenov}}]{2014CoPhC.185..960B}
{B{\'e}langer}, G., {Boudjema}, F., {Pukhov}, A., \& {Semenov}, A. 2014,
  Computer Physics Communications, 185, 960

\bibitem[{{Blasi}(2009)}]{2009PhRvL.103e1104B}
{Blasi}, P. 2009, Physical Review Letters, 103, 051104

\bibitem[{{Boudaud} {et~al.}(2015){Boudaud}, {Aupetit}, {Caroff}, {Putze},
  {Belanger}, {Genolini}, {Goy}, {Poireau}, {Poulin}, {Rosier}, {Salati},
  {Tao}, \& {Vecchi}}]{2015A&A...575A..67B}
{Boudaud}, M., {Aupetit}, S., {Caroff}, S., {et~al.} 2015, \aap, 575, A67

\bibitem[{{Bovy} \& {Tremaine}(2012)}]{2012ApJ...756...89B}
{Bovy}, J. \& {Tremaine}, S. 2012, \apj, 756, 89

\bibitem[{{Casse} {et~al.}(2002){Casse}, {Lemoine}, \&
  {Pelletier}}]{2002PhRvD..65b3002C}
{Casse}, F., {Lemoine}, M., \& {Pelletier}, G. 2002, \prd, 65, 023002

\bibitem[{{Ciafaloni} {et~al.}(2011){Ciafaloni}, {Comelli}, {Riotto}, {Sala},
  {Strumia}, \& {Urbano}}]{2011JCAP...03..019C}
{Ciafaloni}, P., {Comelli}, D., {Riotto}, A., {et~al.} 2011, \jcap, 3, 019

\bibitem[{{Cirelli} {et~al.}(2011){Cirelli}, {Corcella}, {Hektor}, {H{\"u}tsi},
  {Kadastik}, {Panci}, {Raidal}, {Sala}, \& {Strumia}}]{2011JCAP...03..051C}
{Cirelli}, M., {Corcella}, G., {Hektor}, A., {et~al.} 2011, \jcap, 3, 051

\bibitem[{{Coste} {et~al.}(2012){Coste}, {Derome}, {Maurin}, \&
  {Putze}}]{2012A&A...539A..88C}
{Coste}, B., {Derome}, L., {Maurin}, D., \& {Putze}, A. 2012, \aap, 539, A88

\bibitem[{{Delahaye} {et~al.}(2010){Delahaye}, {Lavalle}, {Lineros}, {Donato},
  \& {Fornengo}}]{2010A&A...524A..51D}
{Delahaye}, T., {Lavalle}, J., {Lineros}, R., {Donato}, F., \& {Fornengo}, N.
  2010, \aap, 524, A51

\bibitem[{{Delahaye} {et~al.}(2009){Delahaye}, {Lineros}, {Donato}, {Fornengo},
  {Lavalle}, {Salati}, \& {Taillet}}]{2009A&A...501..821D}
{Delahaye}, T., {Lineros}, R., {Donato}, F., {et~al.} 2009, \aap, 501, 821

\bibitem[{{Delahaye} {et~al.}(2008){Delahaye}, {Lineros}, {Donato}, {Fornengo},
  \& {Salati}}]{2008PhRvD..77f3527D}
{Delahaye}, T., {Lineros}, R., {Donato}, F., {Fornengo}, N., \& {Salati}, P.
  2008, \prd, 77, 063527

\bibitem[{{Di Bernardo} {et~al.}(2013){Di Bernardo}, {Evoli}, {Gaggero},
  {Grasso}, \& {Maccione}}]{2013JCAP...03..036D}
{Di Bernardo}, G., {Evoli}, C., {Gaggero}, D., {Grasso}, D., \& {Maccione}, L.
  2013, \jcap, 3, 36

\bibitem[{{Di Mauro} {et~al.}(2014){Di Mauro}, {Donato}, {Fornengo}, {Lineros},
  \& {Vittino}}]{2014JCAP...04..006D}
{Di Mauro}, M., {Donato}, F., {Fornengo}, N., {Lineros}, R., \& {Vittino}, A.
  2014, \jcap, 4, 6

\bibitem[{{Di Mauro} {et~al.}(2016){Di Mauro}, {Donato}, {Fornengo}, \&
  {Vittino}}]{2016JCAP...05..031D}
{Di Mauro}, M., {Donato}, F., {Fornengo}, N., \& {Vittino}, A. 2016, \jcap, 5,
  031

\bibitem[{{Donato} {et~al.}(2004){Donato}, {Fornengo}, {Maurin}, {Salati}, \&
  {Taillet}}]{2004PhRvD..69f3501D}
{Donato}, F., {Fornengo}, N., {Maurin}, D., {Salati}, P., \& {Taillet}, R.
  2004, \prd, 69, 063501

\bibitem[{{DuVernois} {et~al.}(2001){DuVernois}, {Barwick}, {Beatty},
  {Bhattacharyya}, {Bower}, {Chaput}, {Coutu}, {de Nolfo}, {Lowder}, {McKee},
  {M{\"u}ller}, {Musser}, {Nutter}, {Schneider}, {Swordy}, {Tarl{\'e}},
  {Tomasch}, \& {Torbet}}]{2001ApJ...559..296D}
{DuVernois}, M.~A., {Barwick}, S.~W., {Beatty}, J.~J., {et~al.} 2001, \apj,
  559, 296

\bibitem[{{Ferri{\`e}re}(2001)}]{2001RvMP...73.1031F}
{Ferri{\`e}re}, K.~M. 2001, Reviews of Modern Physics, 73, 1031

\bibitem[{{Fisk}(1971)}]{1971JGR....76..221F}
{Fisk}, L.~A. 1971, \jgr, 76, 221

\bibitem[{Genolini {et~al.}(2015)Genolini, Putze, Salati, \&
  Serpico}]{2015A&A...580A...9G}
Genolini, Y., Putze, A., Salati, P., \& Serpico, P.~D. 2015, Astron.
  Astrophys., 580, A9

\bibitem[{{Genolini} {et~al.}(2016){Genolini}, {Salati}, {Serpico}, \&
  {Taillet}}]{2016arXiv161002010G}
{Genolini}, Y., {Salati}, P., {Serpico}, P., \& {Taillet}, R. 2016, ArXiv
  e-prints [\eprint[arXiv]{1610.02010}]

\bibitem[{{Ghelfi} {et~al.}(2016){Ghelfi}, {Barao}, {Derome}, \&
  {Maurin}}]{2016A&A...591A..94G}
{Ghelfi}, A., {Barao}, F., {Derome}, L., \& {Maurin}, D. 2016, \aap, 591, A94

\bibitem[{{Giesen} {et~al.}(2015){Giesen}, {Boudaud}, {G{\'e}nolini}, {Poulin},
  {Cirelli}, {Salati}, \& {Serpico}}]{2015JCAP...09..023G}
{Giesen}, G., {Boudaud}, M., {G{\'e}nolini}, Y., {et~al.} 2015, \jcap, 9, 023

\bibitem[{{Hooper} {et~al.}(2009){Hooper}, {Blasi}, \& {Dario
  Serpico}}]{2009JCAP...01..025H}
{Hooper}, D., {Blasi}, P., \& {Dario Serpico}, P. 2009, \jcap, 1, 25

\bibitem[{{Kamae} {et~al.}(2006){Kamae}, {Karlsson}, {Mizuno}, {Abe}, \&
  {Koi}}]{2006ApJ...647..692K}
{Kamae}, T., {Karlsson}, N., {Mizuno}, T., {Abe}, T., \& {Koi}, T. 2006, \apj,
  647, 692

\bibitem[{{Kappl} {et~al.}(2015){Kappl}, {Reinert}, \&
  {Winkler}}]{2015JCAP...10..034K}
{Kappl}, R., {Reinert}, A., \& {Winkler}, M.~W. 2015, \jcap, 10, 034

\bibitem[{{Korsmeier} \& {Cuoco}(2016)}]{2016arXiv160706093K}
{Korsmeier}, M. \& {Cuoco}, A. 2016, ArXiv e-prints
  [\eprint[arXiv]{1607.06093}]

\bibitem[{{Lavalle} {et~al.}(2014){Lavalle}, {Maurin}, \&
  {Putze}}]{2014PhRvD..90h1301L}
{Lavalle}, J., {Maurin}, D., \& {Putze}, A. 2014, \prd, 90, 081301

\bibitem[{{Lin} {et~al.}(2015){Lin}, {Yuan}, \& {Bi}}]{2015PhRvD..91f3508L}
{Lin}, S.-J., {Yuan}, Q., \& {Bi}, X.-J. 2015, \prd, 91, 063508

\bibitem[{{Linden} \& {Profumo}(2013)}]{2013ApJ...772...18L}
{Linden}, T. \& {Profumo}, S. 2013, \apj, 772, 18

\bibitem[{{Maurin} {et~al.}(2001){Maurin}, {Donato}, {Taillet}, \&
  {Salati}}]{2001ApJ...555..585M}
{Maurin}, D., {Donato}, F., {Taillet}, R., \& {Salati}, P. 2001, \apj, 555, 585

\bibitem[{{Mertsch} \& {Sarkar}(2014)}]{2014PhRvD..90f1301M}
{Mertsch}, P. \& {Sarkar}, S. 2014, \prd, 90, 061301

\bibitem[{{Moskalenko} \& {Strong}(1998)}]{1998ApJ...493..694M}
{Moskalenko}, I.~V. \& {Strong}, A.~W. 1998, \apj, 493, 694

\bibitem[{{Navarro} {et~al.}(1997){Navarro}, {Frenk}, \&
  {White}}]{1997ApJ...490..493N}
{Navarro}, J.~F., {Frenk}, C.~S., \& {White}, S.~D.~M. 1997, \apj, 490, 493

\bibitem[{{Norbury} \& {Townsend}(2007)}]{2007NIMPB.254..187N}
{Norbury}, J.~W. \& {Townsend}, L.~W. 2007, Nuclear Instruments and Methods in
  Physics Research B, 254, 187

\bibitem[{{Profumo}(2012)}]{2012CEJPh..10....1P}
{Profumo}, S. 2012, Central European Journal of Physics, 10, 1

\bibitem[{{Ptuskin} {et~al.}(1997){Ptuskin}, {Voelk}, {Zirakashvili}, \&
  {Breitschwerdt}}]{1997A&A...321..434P}
{Ptuskin}, V.~S., {Voelk}, H.~J., {Zirakashvili}, V.~N., \& {Breitschwerdt}, D.
  1997, Astronomy and Astrophysics, 321, 434

\bibitem[{{Putze} {et~al.}(2010){Putze}, {Derome}, \&
  {Maurin}}]{2010A&A...516A..66P}
{Putze}, A., {Derome}, L., \& {Maurin}, D. 2010, \aap, 516, A66

\bibitem[{{Strong} \& {Moskalenko}(1998)}]{1998ApJ...509..212S}
{Strong}, A.~W. \& {Moskalenko}, I.~V. 1998, \apj, 509, 212

\bibitem[{{Strong} \& {Moskalenko}(2001)}]{2001AdSpR..27..717S}
{Strong}, A.~W. \& {Moskalenko}, I.~V. 2001, Advances in Space Research, 27,
  717

\bibitem[{{Tan} \& {Ng}(1983)}]{1983JPhG....9.1289T}
{Tan}, L.~C. \& {Ng}, L.~K. 1983, Journal of Physics G Nuclear Physics, 9, 1289

\bibitem[{{Yoon} {et~al.}(2011){Yoon}, {Ahn}, {Allison}, {Bagliesi}, {Beatty},
  {Bigongiari}, {Boyle}, {Childers}, {Conklin}, {Coutu}, {DuVernois}, {Ganel},
  {Han}, {Jeon}, {Kim}, {Lee}, {Lutz}, {Maestro}, {Malinine}, {Marrocchesi},
  {Minnick}, {Mognet}, {Nam}, {Nutter}, {Park}, {Park}, {Seo}, {Sina},
  {Swordy}, {Wakely}, {Wu}, {Yang}, {Zei}, \& {Zinn}}]{2011ApJ...728..122Y}
{Yoon}, Y.~S., {Ahn}, H.~S., {Allison}, P.~S., {et~al.} 2011, \apj, 728, 122

\end{thebibliography}
\end{document}